\documentclass[12pt]{JHEP3}   
\usepackage{amsfonts}   
\usepackage{amsmath}   
\usepackage{multicol}   
\usepackage{epsfig}   
\usepackage{graphicx}   
   
\def\G{\Gamma}    
\def\a{\alpha}  
\def\tq{\tilde {\cal Q}}
   
\title{\Large Super Background Field Method for N=2 SYM}   
\author{Pietro A. Grassi\\ C.N. Yang Institute for Theoretical Physics,  
SUNY at Stony Brook\\ 
Stony Brook, NY 11794-3840, USA}   
\author{Tobias Hurth\\ CERN, Theory Division,  
CH-1211 Geneva 23, Switzerland and \\  SLAC, Stanford University, 
Stanford, CA 94309, USA }   
\author{Andrea Quadri\\ Max-Planck-Institute for Physics  
(Werner-Heisenberg-Institute) \\ 
F\"ohringer Ring 6, D-80805, Munich, Germany} 
   
\abstract{   
The implementation of the Background Field Method (BFM) for quantum field theories 
is analysed within the Batalin--Vilkovisky (BV) formalism.   
We provide a systematic way of constructing general splittings of the fields into classical   
and quantum parts, such that the background 
transformations of the  quantum fields are linear in  
the quantum variables. This leads to linear Ward--Takahashi identities for the background 
invariance  
and to great simplifications in multiloop computations.  
In addition, the gauge fixing is obtained by means  
of (anti)canonical transformations generated by the gauge-fixing  fermion.  
Within this framework we derive the BFM for the N=2 super-Yang--Mills  
theory in the Wess--Zumino gauge viewed   
as the twisted version of Donaldson--Witten topological gauge theory.   
We obtain the background transformations for the full BRST differential  
of N=2 super-Yang--Mills (including gauge transformations,  
SUSY transformations and translations). The BFM permits 
all observables of the supersymmetric theory to be identified easily   
by computing the equivariant cohomology of the topological theory. 
These results should be regarded as  
a step towards the construction  of a  super BFM for the Minimal Supersymmetric Standard Model.}

\preprint{     
YITP-2002-57   \\   
CERN-TH/2002-369 \\   
SLAC-PUB-9853\\
MPI-Pht-2002-81   

}      
     
\begin{document}     
     
\section{Introduction}     
     
One of the most efficient techniques to perform computations in the framework      
of quantum field theory and string theory is the background field method (BFM).      
By  introducing suitable classical background fields in the theory, it is possible      
to derive local Ward--Takahashi identities, which implement the background gauge      
transformations. The latter should be {\it linear} in quantum fields, in contrast      
to the BRST symmetry, which yields non-linear transformations of the      
quantum fields and, correspondingly, the Slavnov--Taylor (ST) identities for the     
quantum effective action.

As a consequence, the Ward--Takahashi (WT) identities for      
the background  gauge invariance relate Green's functions at the same order   
of perturbation theory  
 and they do not require the renormalization of those  composite operators,    
associated to the BRST transformations, which are   
non-linear in the quantized fields.   
     
At the level of  the effective action,     
the background WT identities hold together with the ST ones,     
provided a suitable choice of a background gauge-invariant     
gauge-fixing has been performed.      
It turns out that the Green functions of physical BRST invariant     
operators can be computed by starting from the renormalized background     
gauge-invariant effective action, fulfilling the ST identities,     
after dropping the dependence on      
the quantum fields. The (physical)   
connected functions are then obtained     
by taking the Legendre transform with respect to the background fields,     
once a suitable gauge-fixing for the classical background fields     
is introduced \cite{Abbott:zw,Becchi:1999ir,Ferrari:2000yp}.     
It is this property, together with the advantages provided by the linearity     
of the WT identity in the process of renormalization \cite{Kluberg-Stern:rs,Abbott:1980hw},    
that renders the BFM so appealing.

The fact that the correlation functions of gauge invariant observables   
can be equivalently computed within the BFM technique and with   
the conventional perturbative expansion (together with the conventional   
gauge fixing) can be expressed from the cohomological point of view by   
requiring that the dependence of  Green functions upon the   
background fields be BRST trivial 
\cite{Abbott:zw,Becchi:1999ir,Ferrari:2000yp,Ferrari:2002kz}.     
This can be achieved by enlarging the BRST transformations to 
the background    
fields,  in such a way that they form a set of BRST doublets 
together with their   
corresponding classical background ghosts. As a consequence,   
 the BRST cohomology is unaffected by the presence of those new   
classical fields.   
     
Such a procedure has been applied in   
\cite{Howe:vm,Abbott:1980hw,Denner:nh,Townsend:pg,Grassi:1999nb} to   
the case of models with closed algebras.  In \cite{Howe:vm} the   
(non-linear) splitting of the scalar fields is achieved by using   
normal coordinates on the Riemann manifold, which leads to linear   
background gauge transformations of the quantum fields. On the other   
hand in \cite{skenderis} open gauge algebras within the    
Batalin--Vilkovisky (BV) formalism   
were studied, but the linear splitting of the fields is assumed from   
the beginning, so that  it does not apply directly for example to those   
models where non-linear splittings are necessary   
in order to derive {\em linear} WT identities.  
This is the case for instance of  
N=2 SYM, quantized in the Wess--Zumino gauge, or Donaldson--Witten   
theory, when one wishes to construct the background field transformations   
associated with the full BRST differential (including gauge transformations,  
supersymmetry transformations and translations).  

In the Wess--Zumino gauge the  supersymmetry transformations are non-linear. 
The most convenient way to handle the   
complete set of symmetries \cite{Maggiore,Hollik:1999xh} (gauge   
invariance, supersymmetry, R-symmetry, translations and Lorentz   
transformations) is to construct a generalized BRST operator   
fermionizing the parameters of the rigid symmetries.  
This leads to a set of ST identities, which is difficult to handle and to renormalize.   
For this reason, one would like to construct explicitly the background symmetry for the   
rigid and gauge symmetries of the model. 

This requires a suitable change of variables, by which the original fields of the model   
are split into a background and a quantum part, with the requirement that the new quantum fields    
transform {\it linearly} under the background symmetry.   
If the existence of such a  background symmetry can be established by solving the    
{\it splitting problem}, as we will discuss later,    
the classical action obeys the associated background WT identities.   
   
In the case of supersymmetric theories the conventional BFM can be applied   
to implement background gauge invariance \cite{Kraus:2001id}. However,   
the question of whether it is possible to extend the BFM to the    
full set of rigid and gauge symmetries -- including SUSY transformations,   
R-symmetry, translations and Lorentz transformations -- has to be studied.   
   
At first, we clarify that {\it non-linear} splitting means that the relations   
(splitting functions) between the quantum fields and their backgrounds are    
characterized by complicated expressions involving higher order operators.    
Thus, the non-linear splittings are subject to modifications   
induced by radiative corrections and, consequently, they require new counterterms in   
perturbation theory. However, these functional relations can be constrained by   
symmetry requirements such as background gauge invariance and BRST   
symmetry and, finally, the linearity of the background transformations for    
quantum fields. If these conditions can be solved   
(existence of a solution of the splitting problem) classically,    
the corresponding WTI or STI would bring the same feature   
at the quantum level, namely the number of independent counterterms   
would be  unchanged.    
   
We first study the background splitting problem on 
a general ground 
in the BV formalism  
and we show how    
the antifields could help in the construction of the splitting functions. A by-product of this method   
is the implementation of the BFM by means of canonical transformations that
guarantee    
that the physics of the model is not changed. In addition, also the rigid symmetries can be   
studied in the BV context by promoting the constant parameters of the rigid transformations   
to  be constant ghosts.   
   
In this paper, we briefly analyse the BFM in the Wess--Zumino model    
and present the BFM for the supersymmetry transformations   
within that model. Note that the Wess--Zumino model is an important   
element  in the construction of the MSSM, as it represents both  
matter and  Higgs sectors of the model.  Moreover, 
after the elimination of the auxiliary   
fields, two supersymmetry transformations close on the equations of motion of   
the fermions.  A naive application of the BFM requires an independent background action for the fermions.   
However, the latter is excluded by the invariance under the BRST transformation of the background fields   
Nevertheless, we show how   
this situation can be handled   
within the BV formalism by introducing a further field-antifield pair,  
which is required in order  to take care of the   
closure in the background algebra.

We the consider N=2 SYM    
in the Wess--Zumino (WZ) gauge (in the Euclidean 4-dimensional spacetime).    
In this model the supersymmetry transformation 
of the gaugino is non-linear in the quantum fields, 
a feature  also shared by N=1 SYM in the WZ gauge. 
The latter theory plays a distinguished r\^ole  
since it enters into the construction of the   
Minimal Supersymmetric Standard Model (MSSM) in the WZ gauge 
\cite{Maggiore,Hollik:1999xh},   
where most computations within  MSSM have been carried out.   
 
Having in mind N=1 SYM, we will solve in this paper the splitting 
problem for N=2 SYM, as a first step towards the study of N=1 SYM. 
In the WZ gauge, N=2 SYM  presents some interesting features: the twisted 
formulation is equivalent to  the  
Donaldson--Witten model.
However, in the twisted theory 
the BRST differential has empty cohomology on the   
total space of polynomials in the fields and  antifields. This means   
that it is possible to find a redefinition of fields such that the   
BRST differential can be cast in the form of $s \, {\cal U} = {\cal V}$   
and $s {\cal V} = 0$ of contractible pairs. This simple form allows us   
to construct the linear splitting in the new variables and, mapping   
back to the original variables, the wanted non-linear splitting. The   
background gauge fixing is also studied and both the field   
redefinition and the gauge fixing procedure are achieved by means of   
canonical transformations.   
   
 A legitimate question is how to define the observables in the   
 topological theory, in such a way that they can also be mapped back to   
 the observables of the supersymmetric theory N=2. Following   
 \cite{TYM_sorella}, the observables are defined by computing   
 the BRST cohomology in the space of polynomials with positive powers of the   
 constant ghost $\omega$ (a twisted constant supersymmetric ghost).   However,   
 as pointed out in \cite{Ta-So} the complete cohomology cannot be found in this way,   
 and one has to impose further constraints.   
 The main point is that the correct set of observables is identified in the 
topological version of the   
 theory by means of  
 the equivariant cohomology, as pointed out in \cite{Stora_TYM}. Thus, one has to   
select the space of basic forms   
 on whose space the BRST cohomology is computed. A practical method is to define the basic forms out of the complete space of   
 local polynomials as the kernel of a new nilpotent anticommuting operator $w$ which anti-commutes with the BRST symmetry.   
 The new operator has been constructed in  \cite{Stora_TYM} and it turns out, 
by inspection,  that it generates the   
 background gauge transformations. Therefore, we conclude that observables are selected by   
 computing the BRST cohomology on the space of background-invariant 
operators, which are independent   
 of the background gauge ghost.    
   
   
The paper is organized as follows.  In Section~\ref{geo_BFM},
 we discuss the geometry   
of the splitting and extend the BRST symmetry to the background   
fields; in addition, we provide a general method, based on a cohomological   
analysis, to construct the BFM for a given model. This formulation, 
which relies on the BV formalism, can be applied 
 to implement the BFM for generic models    
with field-dependent and open gauge algebras.     
In Section~\ref{N2SYM}, we apply the construction to the Donaldson--Witten   
model and to N=2 SYM in the Wess--Zumino gauge.    
We construct the BFM for the full BRST differential, thus handling   
in the background formalism the full set of symmetries of the model  
(including gauge symmetries, SUSY transformations and translations).   
In Section~\ref{sec_equiv}, 
the observables for N=2 super-Yang--Mills are defined and    
the appendix contains some auxiliary material.    

\newpage
   
\section{Geometry of BFM}\label{geo_BFM}     
   
\subsection{Symmetries and non-linear splittings}\label{nonlinspl}     
     
We denote by $\xi_i$  the quantum fields and by $\hat \phi_i$ their      
background partners. The original fields of the theory $\Phi_i$ are related to      
$\xi_i$ and  $\hat \phi_i$ by functions      
$\Phi_i =\Phi_i(\hat \phi_i, \xi_i)$. In the following we sometimes use the collective notation      
$\Phi= \{ \Phi_i \}$, $\hat \phi = \{ \phi_i \}$, and  $\xi = \{ \xi_i \}$.      
     
At the classical level, the BRST transformations are described by the rules     
\begin{eqnarray}\label{ns_0}     
s\, \Phi_i = R^\a_i[\Phi(\hat\phi,\xi)] C_\a\,, ~~~~~~~~~~      
s\, C_\a =  {1\over 2} F_\a^{\beta\gamma}[ \Phi(\hat\phi,\xi)] C_\beta C_\gamma\, ,      
\end{eqnarray}     
where $ C_\a$ denote the ghost fields; $R^\a_i[\Phi(\hat\phi,\xi)] $      
are often assumed to be linear functions of the fields $\Phi$.      
This requirement is fulfilled by many gauge theories     
for which the BFM has been implemented, as for instance     
Yang--Mills theory and the Standard Model.     
For the moment we limit ourselves to the case in which $F_\a^{\beta\gamma}$ are constant. 
$F_\a^{\beta\gamma}$ are antisymmetric in the      
$\beta\gamma$ indices, they are related to $R^{\a}_j[\Phi]$ by  the algebra     
\begin{eqnarray}\label{ns_2}     
R^{\a}_j[\Phi] {\delta R^{\beta}_i[\Phi] \over \delta \Phi_j}  -      
R^{\beta}_j[\Phi] {\delta R^{\alpha}_i[\Phi] \over \delta \Phi_j}      
= F_\gamma^{\alpha\beta} R^{\gamma}_i[\Phi] \,,     
\end{eqnarray}     
and satisfy the Jacobi identities $F_\a^{\beta[\gamma} F_\beta^{\delta\sigma]} = 0$.       
In the next subsection we will also consider more general      
situations where $R^\a_i[\Phi(\hat\phi,\xi)] $ and       
$F_\a^{\beta\gamma}[ \Phi(\hat\phi,\xi)]$ are polynomial expressions of the fields, and      
the algebra (\ref{ns_0}) is closed only on-shell.  
     
For the background fields, we assign the following transformation rules     
\begin{eqnarray}\label{ns_3}     
&& s\, \hat\phi_i = \Omega_i + R^\a_i[ \hat\phi] \hat c_\a\,, ~~~~~~~~~~     
      s\, \hat c_\a = \theta_\a  + {1\over 2} F_\a^{\beta\gamma} \hat c_\beta  \, \hat c_\gamma\,, \nonumber \\     
&& s\, \Omega_i = \Omega_j {\delta R^{\a}_i[\hat\phi] \over \delta \hat\phi_j} \hat c_\a -     
                               R^\a_i[\hat \phi] \theta_\a \,,  ~~~~~~~~~~     
      s\, \theta_\a = F_\a^{\beta\gamma} \hat c_\beta \theta_\gamma\,,     
\end{eqnarray}     
where $\hat c_\a$ are the backgrounds for the ghost fields associated to the     
background gauge symmetry.  
The new fields $\Omega_i$ and $ \theta_\a$ are introduced 
in order to control 
the dependence of the theory upon the background fields $\hat \phi_i$ and    
the background   ghosts $\hat c_\a$.
The BRST transformations in eq.(\ref{ns_3}) are nilpotent    
if the functions $R^{\a}_j[\hat\phi]$ are linear. It has been proven (see for example    
\cite{Henneaux:ig,Grassi:1995wr,Quadri:2002nh})    
that the BRST cohomologies $H(s)$ and   $H(s|d)$ are independent of the      
fields $\hat\phi_i, \hat c_\a, \Omega_i$ and $\theta_\a$ and, therefore, the      
physical observables are not affected
by the inclusion of such additional variables.
   
Notice that the structure      
of the BRST transformations for the background fields and the ghosts resemble the      
BRST symmetry for topological models. This observation has been used in      
\cite{Sorella:1989ri} to analyse the BRST cohomology for topological      
sigma models and will play a r\^ole in the forthcoming analysis.  
     
The next step is to split the fields $\Phi_i$ into a quantum and a classical      
part       
\begin{eqnarray}     
\label{split}     
\Phi_i =\hat \phi_i + \Pi_i(\hat\phi, \xi) \, ,     
\end{eqnarray}     
 where $\Pi_i(\hat\phi, \xi) = O(\xi)$,     
in such a way that the background gauge transformations of the      
quantum fields $\xi_i$ are linear in the quantum fields \cite{Howe:vm}.      
This leads to simple linear Ward--Takahashi identities for the Green functions.      
We start with the most general ansatz      
\begin{equation}     
  \label{np_2} s\, \xi_i = P^\a_i( \hat \phi, \xi) C_\a  +      
  S_i^\a( \hat \phi, \xi) \hat c_\a + Q^j_i (\hat \phi, \xi) \Omega_j \,,     
\end{equation}     
where $P^\a_i( \hat \phi, \xi), Q^j_{i} (\hat \phi, \xi)$ and $S_i^\a(\hat \phi, \xi)$      
are 
differential operators depending     
on background and quantum fields.  The linearity condition   
for the background transformation of $\xi_i$  
yields $S_i^\a(\hat \phi, \xi)$ to be linear in $\xi_i$.      
 A linear splitting means that also $\Pi_i(\hat \phi, \xi)$     
in eq.(\ref{split}) is at most linear in $\xi_i$,    
but as we anticipate, this is not always possible.      
     
By inserting the ansatz (\ref{np_2}) in the transformation rules (\ref{ns_0}),      
we derive the following equations     
\begin{eqnarray}     
  \label{ns_4}       
R^\a_k[\Phi(\hat\phi,\xi)] &=& P^\a_i(\hat \phi, \xi) {\delta \Pi_k     
  \over \delta \xi_i} \,, \nonumber \\      
0 &=& \left( \delta^k_{i} + {\delta \Pi_i \over \delta \hat\phi_k} \right) +     
  {\delta \Pi_i  \over \delta \xi_l } Q^k_{l}(\hat \phi, \xi) \,, \nonumber\\     
0 &=&   \left( \delta^k_{i} + {\delta \Pi_i \over \delta \hat\phi_k} \right) R^\a_k( \hat\phi) +      
{\delta  \Pi_i \over \delta \xi_k} S_k^\a( \hat \phi, \xi) \,,     
\end{eqnarray}     
which can be viewed either as consistency conditions for the functions      
$P^\a_i( \hat \phi, \xi), Q^j_{i} (\hat \phi, \xi)$ and $S_i^\a(\hat \phi, \xi)$, given      
the splitting functions $\Pi_i(\hat\phi,\xi)$, or as a constructon tool to      
obtain the splitting, by assuming the transformation rules (\ref{np_2}).    
   
For example,  by eliminating the combination $(1 + {\delta \Pi / \delta \hat\phi})$    
from the second and  the third equation, and by assuming that    
${\delta  \Phi_i / \delta \xi_l}$ is an invertible      
matrix, one has      
\begin{equation}     
  \label{ns_5}     
S^\a_l(\hat\phi,\xi) =  Q_l^k(\hat\phi,\xi) R^\a_k(\hat\phi)\,,      
\end{equation}     
which implies that also the function  $Q_l^k(\hat\phi,\xi)$ is linear in the      
quantum fields $\xi_i$.      
     
In some cases $S_i^\a( \hat \phi, \xi)$ turns out      
to be non-linear in the fields $\xi_i$.    
Moreover, it can happen that the BRST transformation of the ghost fields     
is non-linear in the fields $\xi^i$. In this case it is necessary      
to decompose also the original ghost fields $C^\a$     
in (\ref{ns_0}) into  $C^\a = \hat c^\a + {\cal G}^\a_\beta\, (\hat\phi,\xi) \xi^\beta_C$     
where  $\hat c^\a$ is the background  ghost introduced in (\ref{ns_3}).    
In order to  respect the ghost number, the function ${\cal G}^\a_\beta \, (\hat\phi,\xi)$       
depends on zero-ghost number fields. Equation~(\ref{np_2}) now reads     
\begin{equation}      
\label{np_3}      
s\, \xi_i = P^\a_i( \hat \phi, \xi)  {\cal G}_{\a\beta}(\hat\phi,\xi) 
\xi_C^\beta +      
{\cal S}_i^\a( \hat \phi, \xi) \hat c_\a + Q^j_i(\hat \phi, \xi) \Omega_j \,,     
\end{equation}     
where ${\cal S}_i^\a( \hat \phi, \xi) = P^\a_i( \hat \phi, \xi) + S_i^\a( \hat \phi, \xi)$.      
The splitting of the ghost fields is chosen in such a way that ${\cal S}_i^\a( \hat \phi, \xi)$      
is a linear function of the quantum fields, namely of $\xi$.  
     
For Yang--Mills theory \cite{Kluberg-Stern:rs}, we can identify the      
symbols in eqs.~(\ref{ns_4}) and~(\ref{np_3}) with the conventional notation:    
$\hat\phi_i \equiv \hat A^\a_\mu$,      
$\xi_i \equiv Q^\a_\mu$ , $R^\a_k(\Phi) \equiv \nabla^a_\mu$ and      
$Q_l^k(\hat\phi,\xi) \equiv \delta^a_b \delta^\mu_\nu$. It is easy to see      
that $\Pi_\mu^a(\hat A,Q) = Q_\mu^a$ solves eq.~(\ref{ns_4}) and eq.~(\ref{ns_5}). Notice      
that there is a more general solution to eqs.~(\ref{ns_4}) with $\Pi_\mu^a(\hat A,Q)      
=\Theta^{ab}_{\mu\nu}(\hat A) Q^\nu_b$,      
where $ \Theta^{ab}_{\mu\nu}(\hat A)$ is a combination      
of background gauge invariant operators. From equation~(\ref{ns_3}) we see that      
\begin{eqnarray}     
s\hat A_\mu^a = \Omega_\mu^a + \partial_\mu \hat c^a + f^{abc} \hat A_\mu^b \hat c^c \, .     
\label{q1}     
\end{eqnarray}     
Then from eq.~(\ref{ns_0}) we obtain     
\begin{eqnarray}     
s Q_\mu^a =  \partial_\mu (C^a - \hat c^a) + f^{abc}\hat A_\mu^b (C^c - \hat c^c)     
+ f^{abc} Q_\mu^b C^c - \Omega_\mu^a \,.     
\label{q2}     
\end{eqnarray}     
By splitting the original ghost  $C^a$ into $C^a = \hat c^a + \xi_C^a$ we can rewrite (\ref{q2}) in the      
following way     
\begin{eqnarray}     
s Q_\mu^a = \partial_\mu \xi_C^a + f^{abc}\hat A_\mu^b  \xi_C^c + f^{abc} Q_\mu^b \xi_C^c 
+ f^{abc} Q_\mu^b \hat c^c - \Omega_\mu^a \, .     
\label{q3}     
\end{eqnarray}

The background transformation of $Q_\mu^a$ has to be identified with     
the fourth term in the above equation, which is of order 1      
in the ghost background $\hat c^a$. This leads to the identification      
${\cal S}^\a_i(\hat\phi,\xi) \equiv  f^{abc} Q_\mu^b$. Moreover,      
the third  term is bilinear in the quantum fields and      
$P^\a_i( \hat \phi, \xi)  {\cal G}_{\a\beta} (\hat\phi,\xi)\equiv      
\hat\nabla^{ab}_\mu  - f^{abc} Q_\mu^c$.      
A suitable choice of the splitting of ghost fields simplifies the      
structure of quantum gauge transformations.      
     
For non-linear sigma models (cf.~\cite{Howe:vm}), the gauge transformations      
(\ref{ns_0}) for the coordinates $\Phi_i$ of the manifold are replaced by      
diffeomorphisms $R^\a_i(\Phi) C_\a \rightarrow  v_i(\Phi)$, where $v_i$ are the components of      
a vector field\footnote{The transformations $ \delta \Phi_i = v_i(\Phi)$ are rigid transformations      
from the worldsheet point of view. They can be translated into BRST transformations      
by decomposing $v(\Phi)$ into power series and fermionizing the      
coefficients: $v(\Phi) = \sum_n v_n \Phi^n \rightarrow \sum_n c_n \Phi^n$ where      
$c_n$ are an infinite set of constant anticommuting ghosts. Then, we have $s\, \Phi_i  = \sum_n c_n \Phi^n$ and      
$ s\, c_n = \sum_m \, (m-n) c_{m-n } c_n$. The latter are the usual BRST transformations      
for the ghost fields for the Virasoro algebra.}      
and eqs.~(\ref{ns_4}) are rewritten in the form     
\begin{eqnarray}     
  \label{ns_6}     
v_i(\Phi) &=&\left(\delta^k_i + {\delta \Pi_i \over \delta \hat\phi_k }\right)  v_k(\hat\phi) +      
\xi_l \partial^l v_k(\hat\phi) {\delta \Pi_i \over \delta \xi_k} \nonumber \\     
0 &=&      
 \left( \delta^k_{i} + {\delta \Pi_i \over \delta \hat\phi_k} \right) +     
  {\delta \Pi_i  \over \delta \xi_l } Q^k_{l}(\hat \phi, \xi) \,,      
\end{eqnarray}     
and    
$S_i(\hat\phi,\xi) =  \xi_k \partial^k v_i(\hat\phi)$.\footnote{We remind the reader that     
if $\xi$ is a vector field, it is natural to define $\delta \xi = {\cal L}_v \xi$, where ${\cal L}_v$ is the      
Lie derivative. This means that $\delta (\xi_i \partial^i) = (\xi_k \partial^k v_i - v_k \partial^k \xi_i) \partial^i$.      
Assuming that $\xi_i$ are constant and independent of $\hat\phi_i$, we have      
$\delta \xi_i = \xi_k \partial^k v_i$.} Following \cite{Howe:vm}, we can use an      
interpolating field $\Phi_i(t)$,  which satisfies the geodesic equations for a given      
connection $\G^i_{jk}$ and construct a solution to (\ref{ns_6})     
\begin{eqnarray}     
&&\Phi_i(\hat\phi,\xi) = \hat\phi_i + \Pi_i(\hat\phi, \xi)\,, ~~~~~~~      
\Pi_i(\hat\phi, \xi) = \xi_i + \chi_i(\hat\phi,\xi) \,, \nonumber      
\\     
&&\chi_i (\hat\phi,\xi)= -\sum_{n=2}^\infty \frac{1}{n!} \G_i^{j_1\dots j_n}(\hat\phi) \xi_{j_1} \dots \xi_{j_n} \,,     
\label{geo8}     
\end{eqnarray}     
where $\G_i^{j_1\dots j_n}(\hat\phi)$ are related to the covariant derivatives    
of the connection $\G^i_{jk}$ computed at the point $\hat \phi_i$.     
In this example we can easily justify the invertibility of the matrix    
$\frac{\delta \Phi_i}{\delta \xi_j}$;   in fact, we have     
\begin{eqnarray}     
\frac{\delta \Phi_i}{\delta \xi_j} = \delta_i^j + O(\xi) \,     
\label{geo9}     
\end{eqnarray}     
and, by equations.~(\ref{geo8}), $\delta \Phi_i / \delta \xi_j $ is invertible as a formal power series.     
Equation~(\ref{geo8}) provides an explicit and particular      
solution for the splitting. This is not the only possible solution      
compatible with a linear transformation of the quantum field      
${\cal S}_i(\hat\phi,\xi) =  \xi_k \partial^k v_i(\hat\phi)$.      
However we point out that~(\ref{geo8}) turns out to be the      
most general solution (up to rotations in the $\xi$-space)     
for the background transformation of the quantum field.     
     
Following these suggestions of the non-linear sigma model,    
we can deduce that the most general solution      
for ${\cal S}^\a_i(\hat\phi,\xi)$ is given by      
\begin{eqnarray}     
{\cal S}_i^\a (\hat \phi,\xi) = \frac{\delta R^\a_i(\hat \phi)}{\delta \hat \phi_j} \xi_j \,.     
\label{geo4}     
\end{eqnarray}     
To prove this assertion, we insert equation~(\ref{split}) into       
the third of equations~(\ref{ns_4}) and we expand up to the first order in $\xi_j$. We get     
\begin{eqnarray}     
&& R^\a_k[\hat\phi]      
+ \frac{\delta R^\a_k[\hat\phi]}{\delta \hat \phi_j}     
  \left . \frac{\delta \Pi_j}{\delta  \xi_p} \right |_{\xi=0} \xi_p + \dots      
 =       
R^\a_k[\hat\phi]      
+ \left . \frac{\delta^2 \Pi_k}{\delta \hat \phi_i \delta \xi_p} \right|_{\xi=0}     
  \xi_p R_i^\a[\hat\phi] \nonumber \\     
&& ~~~~~~~~ +  \left . \frac{\delta \Pi_k}{\delta  \xi_j} \right |_{\xi=0}     
       \left . S^\a_j \right |_{\xi=0}     
    + \left . \frac{\delta \Pi_k}{\delta  \xi_j} \right |_{\xi=0}      
      \left . \frac{\delta  S^\a_j}{\delta \xi_p} \right |_{\xi=0} \xi_p      
    + \left . \frac{\delta^2 \Pi_k}{\delta \xi_i \delta \xi_p} \right |_{\xi=0}      
      \xi_p \left . S^\a_i \right |_{\xi=0} + \dots  \nonumber \\    
\label{geo12}    
\end{eqnarray}     
By looking at the terms of order zero in $\xi$ we get     
\begin{eqnarray}     
 R^\a_k[\hat\phi] =  R^\a_k[\hat\phi]  +       
\left . \frac{\delta \Pi_k}{\delta  \xi_j} \right |_{\xi=0}     
       \left . S^\a_j \right |_{\xi=0} \, .     
\label{geo13}     
\end{eqnarray}     
This gives $\left . S^\a_j \right |_{\xi=0} = 0$ since    
$\left . \frac{\delta \Pi_k}{\delta  \xi_j} \right |_{\xi=0}$ is invertible.     
Then we look at the terms of order one in $\xi$     
in equation~(\ref{geo12}), by taking into account $\left . S^\a_j \right |_{\xi=0} = 0$ and      
the fact that $\left . \frac{\delta^2 \Pi_k}{\delta \hat \phi_i \delta \xi_p} \right|_{\xi=0} = 0$      
because of the invertibility of $\frac{\delta \Pi_k}{\delta  \xi_j}$ as a formal power series.     
We obtain      
\begin{eqnarray}     
 \frac{\delta R^\a_k[\hat\phi]}{\delta \hat \phi_j}     
  \left . \frac{\delta \Pi_j}{\delta  \xi_p} \right |_{\xi=0} =          
 \left . \frac{\delta \Pi_k}{\delta  \xi_j} \right |_{\xi=0}      
      \left . \frac{\delta  S^\a_j}{\delta \xi_p} \right |_{\xi=0} \,,     
\label{geo16}     
\end{eqnarray}     
and finally     
\begin{eqnarray}     
 \left . \frac{\delta  S^\a_i}{\delta \xi_p} \right |_{\xi=0}      
=  \left . \frac{\delta  \xi_i}{\delta \Pi_k} \right |_{\xi=0}      
  \frac{\delta R^\a_k[\hat\phi]}{\delta \hat \phi_j}     
  \left . \frac{\delta \Pi_j}{\delta  \xi_p} \right |_{\xi=0} \, .     
\label{geo17}     
\end{eqnarray}     
This means that up to a rotation in the $\xi$-space we recover equation~(\ref{geo4}).     
    
\subsection{BV formulation of the splitting problem}    
     
The main question is how to solve equations~(\ref{ns_4}) or equations~(\ref{ns_6}) to find      
the splitting for the fields $\Phi_i$ and for the ghost fields $C^\a$.

For this purpose, we rewrite equations~(\ref{ns_4}) in a different form, suitable for the      
direct application of the BV formalism and for a cohomological reformulation      
of the splitting problem. 
We work out the necessary formalism  for the general case of an open algebra.      
    It is convenient to introduce the antifields for each field      
of the model and to modify the classical gauge invariant action $S_0$ into $S = S_0 + S_*$, where      
$S_*$ is given by       
\begin{eqnarray}     
\label{gip1}     
S_* = \int \left(     
\Phi^{*,i} s\, \Phi_i  + C^{*,\a} s\, C_\a +     
\hat\phi^{*,i} s\, \hat \phi_i + \hat c^{*,\a} s\, \hat c_\a +     
\Omega^{*,i} s\, \Omega_i  +  \theta^{*,\a} s\, \theta_\a      
\right) \, .     
\end{eqnarray}     
$S$ obeys the master equation     
\begin{eqnarray}     
(S,S) = 0 \, .     
\label{qn1_I}     
\end{eqnarray}     
The bracket in equation~(\ref{qn1_I}) is defined by     
\begin{eqnarray}     
(X,Y) & \equiv & \int \, \left (     
      \frac{\delta_r X}{\delta \varphi^I} \frac{\delta_l Y}{\delta \varphi^*_I}     
    - \frac{\delta_r X}{\delta \varphi^*_I} \frac{\delta_l Y}{\delta \varphi^I}       \right )     
 \, ,     
\label{qn2_I}     
\end{eqnarray}     
where $\varphi^I = \{ \Phi^i, C^\alpha, \hat \phi^i, \hat c_\a , \Omega_i, \theta_\a\}$,      
$\varphi^*_I = \{ \Phi^{*,i}, C^{*,\alpha}, \hat \phi^{*,i}, \hat c^{*,\a}, \Omega^{*,i}, \theta^{*,\a} \}$.     
In principle, one should not need antifields for classical fields     
such as the background fields $\hat\phi_i$ and for their     
shifts $\Omega_i$, but it turns out that they are needed in order to     
handle open algebras. Indeed, with antifields, one can easily accommodate     
general gauge algebras of the form     
\begin{eqnarray}\label{op_1}     
R^{\a}_j {\delta R^{\beta}_i \over \delta \Phi_j}  -      
R^{\beta}_j {\delta R^{\alpha}_i \over \delta \Phi_j}      
= F_\gamma^{\alpha\beta}[\Phi] R^{\gamma}_i + M^{\alpha\beta}_{ij}[\Phi]     
{\delta S_0 \over \delta \Phi_j} \,,     
\end{eqnarray}     
where $ F_\gamma^{\alpha\beta}[\Phi]$ and     
$M^{\alpha\beta}_{ij}[\Phi]$ involve dynamical variables $\Phi_i$.     
The algebra described by the generators $R^{\a}_j[\Phi]$ is an open algebra     
and the last term in (\ref{op_1}) takes into account those symmetries     
which are closed on the classical equations of motion ${\delta S_0 / \delta \Phi_j} =0$.     
The latter term is not there in the case of a closed algebra of course. 
By consistency with the invariance of action $S_0$, one finds that     
$M^{\alpha\beta}_{ij}[\Phi]  = - M^{\alpha\beta}_{ji}[\Phi]$ and     
$ M^{\alpha\beta}_{ij}[\Phi]  = - M^{\beta\alpha}_{ij}[\Phi]$.     
    
By using the antifields, the BRST transformations are modified into     
\begin{eqnarray}\label{op_2}    
s\, \Phi_i = R^{\a}_i [\Phi] \, C_\a + M^{\beta\alpha}_{ij}[\Phi] \, C_\a C_\beta \Phi^{*,j}\,,    
~~~~~~~    
s\, \Phi^{*,i} = {\delta S\over \delta \Phi_i}\,.     
\end{eqnarray}    
The fulfilment of the master equation in equation~(\ref{qn1_I})      
requires that the action be changed by adding new terms     
that are quadratic in the antifields     
\begin{equation}    
S \rightarrow S + {1\over 2} \int     
M^{\beta\alpha}_{ij}[\Phi] C_\a C_\beta \Phi^{*,i} \Phi^{*,j}\,.    
\end{equation}     
The nilpotency of the BRST transformation on the antifield $\Phi^{*,j}$     
then follows from the invariance of the action $S_0$ and of the antifield     
terms $S_*$.     
    
In more general cases, for     
example in case of reducible gauge theories, one usually needs new terms with     
higher powers of antifields and new ghost fields to parametrize the     
new symmetries.    

\medskip

We notice that, corresponding to the symmetry (\ref{op_2}), we can introduce a background gauge symmetry,     
where we replace $\Phi$ and $\Phi^*$ with the background partners     
everywhere and the ghost $C_\a$ with the background ghost $\hat c_\a$.     
In addition, we still have to add the shift fields generated by $\Omega_i$ and $\theta_\a$.  
At first one might think that the natural definition of      
the background symmetry for the background fields is:
\begin{eqnarray}\label{op_3}    
s\, \hat\phi_i = R^{\a}_i [\hat\phi] \, \hat c_\a + M^{\beta\alpha}_{ij}[\hat\phi] \, \hat c_\a \hat c_\beta     
\hat\phi^{*,j}  + \Omega_i \,,~~~~~    
s\, \hat c_\a = \frac{1}{2}    
F_\a^{\beta\gamma}[\hat\phi] \hat c_\beta \hat c_\gamma     
+ \theta_\a\,.    
\end{eqnarray}     
     
However, this would lead to a difficulty: 
in order to reproduce, for the background fields, the same    
gauge algebra as in equation~(\ref{op_2}), we should add to the classical action    
$S$ new terms in order to generate the ``closure terms,,     
proportional to the equations of motion. This is    
excluded by the presence of the shifts generated by the fields 
$\Omega_i$ and $\theta_\a$. In order to circumvent this problem we     
introduce new antifields, denoted by $\hat\chi^{*,j}$, which      
replace the antifields $\hat\phi^{*,j}$ in equations~(\ref{op_3})    
\begin{eqnarray}\label{op_3_bis}    
s\, \hat\phi_i = R^{\a}_i [\hat\phi] \, \hat c_\a + M^{\beta\alpha}_{ij}[\hat\phi] \,     
\hat c_\a \hat c_\beta     
\hat\chi^{*,j}  + \Omega_i \,.~~~~~    
s\, \hat c_\a = F_\gamma^{\alpha\beta}[\hat\phi] R^{\gamma}_i [\hat\phi] + \theta_\a\,.    
\end{eqnarray}     
Their transformation rules reproduce the correct algebra by imposing    
\begin{eqnarray}\label{op_4.1}    
s\, \hat\chi^{*,i} = \left({\delta S_0[\Phi] \over \delta \Phi_i}\right)_{\Phi \rightarrow \hat\phi}     
+ \Omega^{*,i}_\chi + \dots \,,    
\end{eqnarray}     
where $ \left({\delta S_0 / \delta \Phi_i}\right)_{\Phi \rightarrow \hat\phi} $ is the classical gauge     
covariant equations of motion where the original fields $\Phi_i$ have been replaced by the     
background fields $\hat\phi_i$, $ \Omega^{*,i}_\chi$ are the shift fields for $\hat\chi^{*,i}$    
and the ellipsis denotes further terms with at least one power of antifields,    
eventually needed to guarantee the closure of     
the algebra.   

The requirement of nilpotency of $s$ on $\hat \phi_i$ imposes a constraint on $s\Omega_i$, while nilpotency
on $\hat \chi^{*,i}$ yields a constraint for $s \Omega^{*,i}_\chi$.

Finally, we can introduce the BRST transformations for the     
quantum fields $\xi_i$ and their antifields $\xi^*_i$.     
The only assumption we have to impose here  is that the quantum fields should transform linearly     
under the background gauge transformations     
\begin{eqnarray}\label{opp_1}    
s\, \xi_k &=& \xi_l \left. {\delta R^{\a}_k\over \delta \Phi_l}\right|_{\Phi \rightarrow \hat\phi} \hat c_\a +     
{1\over 2} \hat c_\a \hat c_\beta N^{\a\beta}_{kj}[\hat\phi] \, \xi^{*,j}  + \dots \nonumber \\    
s\, \xi^{*,k} &=& \xi_l  \left. {\delta S_0 \over \delta \Phi_l \delta \Phi_k }\right|_{\Phi \rightarrow \hat\phi} + \dots     
\end{eqnarray}    
where the ellipses denote the BRST transformations generated by the     
quantum ghosts and by the shifts $\Omega_i$. In order to compute these remaining     
terms of rules (\ref{op_4.1}) and (\ref{opp_1}), one has     
to solve the master equation (\ref{qn1_I}) with the bracket  given in terms of
$$\varphi^I = \{ \xi^i, \xi^\alpha_C, \hat \phi^i, \hat c_\a , \Omega_i, \theta_\a,\hat \chi_i, \Omega_{\chi^i}  \}, ~~~    
\varphi^*_I = \{ \xi^{*,i}, \xi^{*,\alpha}_C, \hat \phi^{*,i}, \hat c^{*,\a}, \Omega^{*,i}, \theta^{*,\a}, \hat \chi^{*,i},
\Omega^{*,i}_\chi \}$$
($\xi^\alpha_C$ and $\xi^{*,\alpha}_C$ are the quantum ghost fields and their ``quantum''  antifields, respectively),     
with boundary conditions (\ref{op_3_bis}), (\ref{op_4.1})  and (\ref{opp_1}). It is easy to check that in the     
case of the Yang--Mills theory and the non-linear sigma model, this leads to the usual     
splitting between quantum and classical fields. In more general cases, one has     
to show that there is at least a solution.    

Notice that since all the fields and antifields in the background sector are classical     
fields, there is no distinction between a field and an antifield    
from the point of view of the quantization of the model.     
$\Omega^{*,i}$ removes the reducibility between the ``quantum,, antifield     
$\xi^{*,i}$ and the classical antifield $ \hat\phi^{*,i}$.    
   
\medskip

As an illustration of the previous considerations,  we analyse a simple model, namely    
the N=1 Wess--Zumino model.  There,     
the role of the antifields  $ \hat\chi^{*,i} $ and their shifts    
 $\Omega^{*,i}_\chi$ will become clear. In this case the antifields $ \hat\chi^{*,i} $    
can be interpreted as the background counterpart of the auxiliary field    
$F$.    The same thechnique has been used in \cite{Maggiore}. 
    
The model is written in terms of the fields $\Phi_i = \{ A, \psi^\a \}$,    
where $A$ is a complex scalar field and     
$\psi^\a$ is a Weyl spinor.\footnote{The motivation to consider a BFM formulation for the     
Wess--Zumino models is related to the implementation of the BFM for the MSSM.  
There two Wess--Zumino models for the Higgs superfields  $H_1$ and $H_2$ are coupled to the     
gauge invariant action in order to break the $SU(2)_L \times U_Y(1)$ down to the subgroup     
$U_Q(1)$. In order to write a generalization of the `t Hooft-background gauge fixing for     
the MSSM, one needs  to add the background fields for the scalar components of $H_1$ and $H_2$.     
In addition, in order to mantain the supersymmetry manifest, one has to add also the background     
for their superpartners.} We also introduce the ghost fields $\eta^\a$ for     
the supersymmetry and $v^\mu$ 
for the translations. Since there is no gauge symmetry,     
we consider only rigid supersymmetry transformations. By eliminating the auxiliary fields,    
the algebra of supersymmetry  closes only on-shell. 
The problem can be reformulated in the     
context of the BV framework. The classical action $S = S_0 + S_1 + S_2$ reads    
\begin{eqnarray}\label{op_4}    
S_0 &= & \int d^4x \left( |\partial_\mu A|^2 - i \psi^\a \sigma^\mu_{\a\dot\beta}    
\partial_\mu \bar\psi^{\dot\beta}\right) \,,    
\nonumber \\    
S_1 &=& \int d^4x \left( A^* (s A) + \psi^{*\a} (s \psi_\a) + v^{*\mu} (s v_\mu) + { \rm c.c.} \right) \,,     
\nonumber \\    
S_2 & =& \int d^4x \left( 2 \eta^\a \psi^{*}_\a \, \bar\eta_{\dot\beta} \bar\psi^{*\dot\beta} \right)\,,    
\end{eqnarray}     
where the BRST transformations are given by     
\begin{eqnarray}\label{op_5}    
s\, A &=& 2 \eta^\a \psi_\a - i v^\mu \partial_\mu A\,, ~~~~~~    
s\, \psi_\a  = - i \sigma^\mu_{\a\dot\beta} \bar\eta^{\dot \beta} \partial_\mu A      
- i v^\mu \partial_\mu   \psi_\a   - 2 \eta^\a \, \bar\eta_{\dot\beta} \bar\psi^{*\dot\beta}     
\,, \nonumber\\    
s\, v^\mu &=& - 2 \eta^\a \sigma^\mu_{\alpha \dot\beta} \bar\eta^{\dot \beta}\,,  ~~~~~~~~~~~~    
s\, \eta^\a = 0\,.    
\end{eqnarray}     
The BRST transformation of the fermion $ \psi_\a $ contains the antifield $\bar\psi^{*\dot\beta} $ in order     
to take into account the closure on the equations of motion. 
This is reflected at the level of the classical action in the term $S_2$, quadratic in the antifields.
We can derive the BRST transformations for the background field as     
\begin{eqnarray}\label{op_6}    
s\, \hat A &=& 2 \hat\eta^\a \hat\psi_\a - i \hat v^\mu \partial_\mu \hat A + \Omega_A\,, ~~~~~~\nonumber \\    
s\, \hat\psi_\a  &=& - i \sigma^\mu_{\a\dot\beta} \hat{\bar\eta}^{\dot \beta} \partial_\mu \hat A      
- i \hat v^\mu \partial_\mu  \hat\psi_\a   - 2 \hat\eta_\a \,\hat F    
+ \Omega_{\psi \, \a}
\,, \nonumber\\    
s\, \hat v^\mu &=& - 2 \hat\eta^\a \sigma^\mu_{\alpha \dot\beta}  \hat{\bar\eta}^{\dot \beta} + \theta^\mu_v\,,     
 ~~~~~~~~~~~~ \nonumber \\    
s\, \hat\eta^\a &=& \theta^\a\,.    
\end{eqnarray}     
The fields $\Omega_A$ and $\Omega^\a_\psi$ are the shift for the background fields     
$\hat A$ and $\hat\psi_\a$ and     
the fields $\theta^\a$ and $\theta^\mu_v$ are the background fields for the ghost fields     
$\hat \eta_\a$ and $\hat v^\mu$.     
In the above equation we have reintroduced the auxiliary field $\hat F$.
The BRST transformation rules are given by
\begin{eqnarray}\label{op_7}    
s\, \hat F = - i \hat{\bar\eta}^{\dot\beta} {\bar \sigma}_{\dot\beta \a}^\mu \partial_\mu \hat\psi^\a -     
i \hat v^\mu \partial_\mu \hat F + \Omega_F\,,     
\end{eqnarray}     
where $\Omega_F$ is the corresponding shift field. A simple exercise shows that,     
in order to mantain the nilpotency on $\hat \psi_\a$, one needs to impose    
the following transformation:     
\begin{eqnarray}\label{op_8}    
s \, \Omega_{\psi \, \a} & = & 2 \hat \eta_\a \Omega_F + i     
\hat v^\mu \partial_\mu \Omega_{\psi \, \a} + i \sigma^\mu_{\alpha \dot \beta}     
\hat{ \bar \eta}^{\dot \beta} \partial_\mu \bar \Omega_A \nonumber \\    
&&     
+ 2 \theta_\a \hat F + i \theta^\mu_v \partial_\mu \hat \psi_\a +     
i \sigma^\mu_{\alpha \dot \beta} \bar \theta^{\dot \beta} \partial_\mu \hat A \, ,    
\end{eqnarray}    
which represents the supersymmetry algebra at the level of the     
fields $\Omega^\a_\psi, \Omega_F, \Omega_A$.     
They form a chiral multiplet.\footnote{Equations~(\ref{op_7}) and (\ref{op_8})     
can be obtained in a straightforward way by using a     
superspace technique: $s\, \hat X \equiv \eta^\a D_\a \hat X + v^\mu \partial_\mu \hat X + \Omega$     
where $X$ and $\Omega$ are chiral superfields and $D_\a $ is the covariant derivative.     
By imposing the nilpotency, $s^2=0$, one gets    
$s\, \Omega = \eta^\a D_\a \Omega + v^\mu \partial_\mu \Omega$.     
This means that the fields $\Omega$ transform under the supersymmetry transformations as a     
chiral supermultiplet.}

We can avoid the introduction of the auxiliary fields by using      
an additional antifield, as outlined before. In order to distinguish the     
antifield $\hat\psi^{*,\a}$ (coupled to the BRST variation    
of $\psi^\a$) from the new antifield,     
which is needed to reproduce the correct algebra at the level of the    
background fields,    
 we will denote the latter by $\hat\chi^{*,\a}$.     
    
The BRST transformation for the spinor $\psi_\a$      
is correspondingly given by     
\begin{eqnarray}\label{op_9}    
s\, \hat\psi_\a  &=& - i \sigma^\mu_{\a\dot\beta} \hat{\bar\eta}^{\dot \beta} \partial_\mu \hat{A}      
- i \hat v^\mu \partial_\mu  \hat\psi_\a   - 2 \hat\eta_\a \, \hat{\bar\eta}^{\dot\beta}  \hat\chi^{*}_{\dot\beta}    
+ \Omega_{\psi \, \a}\,.    
\end{eqnarray}    
By requiring the nilpotency of the BRST transformation, we find     
\begin{eqnarray}\label{op_10}    
s\,  \hat\chi^{*}_{\dot\a} = - i     
{\bar \sigma}_{\dot\a \beta}^\mu \partial_\mu \hat\psi^\beta -     
i \hat v^\mu \partial_\mu \hat\chi^{*}_{\dot\a}  + \Omega^*_{\chi,\dot \a}\,.    
\end{eqnarray}    
$\Omega^*_{\chi,\dot \a}$ is the shift for $\hat\chi^{*}_{\dot\a}$     
and it guarantees     
that the cohomology is independent of the variables    
$\Omega^*_{\chi,\dot\alpha}$ and $\hat \chi^*_{\dot\alpha}$.     
The BRST transformations for $\Omega_{\psi \a}$ and    
$\Omega^*_{\chi, \dot\a}$ can again be derived by imposing the nilpotency    
of $s$ on $\hat \psi_\alpha$ and $\hat \chi^*_{\dot\alpha}$.    
    
These BRST transformations (\ref{op_9}) and (\ref{op_10})     
can be implemented within the BV formalism by coupling     
$s\, \hat\psi_\a $ and $s\,  \hat\chi^{*}_{\dot\a}$ to     
the  corresponding conjugate variables (``antifields,,)     
$ \hat\psi^*_\a$ and $ \hat\chi_{\dot\a}$.    
We notice that in the case of $\hat\chi^{*}_{\dot\a}$ the ``antifield,,     
is actually an external source with its own shift field $\Omega_{\chi,\a}$.     
    
This example shows that we have to introduce the antifields $\hat \chi^*$ (or eventually the auxiliary fields)     
for each background field $\hat \phi$ 
on which the BRST differential squares to zero only modulo the equations
of motion. 
This is needed  in order to reproduce the correct algebra. In addition, each     
new antifield $\hat \chi^*$ has to be paired 
with a corresponding shift $\Omega^*_\chi$ in order     
to enforce the triviality with respect to the BRST cohomology and to     
close the symmetry on the antifields $\hat\chi^*$.      

We also remark that by using antifields instead of auxiliary fields,     
we loose the multiplet structure.  It seems that even in the cases where the  
auxiliary fields cannot be found in order to establish the closure of the algebra at the level  
of fields, the BV technique by means of antifields is able to supply the correct content  
of variables to close the algebra. However, the structure of superfields is no longer available.  

\medskip
    
We can finally go back to the initial question how to define the     
correct splitting between the quantum and the classical fields.     
The master equation (\ref{qn1_I}) has to be solved 
in the appropriate space of variables (including the needed auxiliary
antifields)
$$\varphi^I = \{ \xi^i, \xi^\alpha_C, \hat \phi^i, \hat c_\a , \Omega_i, \theta_\a,\hat \chi_i, \Omega_{\chi^i}  \} , ~~~~ \varphi^*_I = \{ \xi^{*,i}, \xi^{*,\alpha}_C, \hat \phi^{*,i}, \hat c^{*,\a}, \Omega^{*,i}, \theta^{*,\a}, \hat \chi^{*,i},
\Omega^{*,i}_\chi \}$$     
with boundary conditions (\ref{op_3_bis})-(\ref{opp_1}) and under the
requirement that
\begin{eqnarray}
\left . S 
\right |_{\varphi^*_I = 0, \xi^i = \xi^\alpha_C = \Omega_i= \theta_\a = \hat \chi_i =\Omega_{\chi^i} =0} =S_0[\hat\phi,\hat c] \, .
\label{addq1}
\end{eqnarray}
$S_0$ is the original gauge invariant 
classical action. We will discuss in the next
section the problems related with the background gauge-fixing.

In the case of closed algebras no auxiliary antifields
are needed and condition (\ref{addq1}) is
fulfilled since the implementation of the BFM yields the replacement
of the original fields $\Phi_i,C^\a$ with
\begin{eqnarray}
\Phi_i = \hat \phi_i + \Pi_{\phi,i} \, , ~~~~ 
C^\a = \hat c^\a + \Pi_{C,\a} \, ,
\label{addq2}
\end{eqnarray}
where $\Pi_{\phi,i}, \Pi_{C,\a}$ are functions
of $\hat \phi_i, \hat C^\a, \xi_i,\xi^\a_C$ vanishing for
$\xi_i=\xi^\a_C=0$.

For open gauge algebras the implementation of the BFM
requires the extension of the space of variables in which the splitting
problem can be defined (due to the introduction of the auxiliary
fields). Equation~(\ref{addq1}) then provides the relation 
of the full classical action $S$ with the
original classical action $S_0$.

The methods needed to solve this problem vary with the model at hand.    
It may happen that a suitable choice of the generators of the original BRST    
differential $s$ 
is enough to obtain a solution. 
This is the case for instance of the Topological    
Yang--Mills (TYM) theory, where the Jordan form of the BRST differential    
\cite{henn_dual} can be reached by a suitable field redefinition.    
Notice that this is only possible if one introduces the relevant set
of auxiliary fields (corresponding to the twisted auxiliary fields
of $N=2$ SYM in the WZ gauge). The use of the auxiliary fields
reduces the open algebra problem to a closed algebra problem.
In the space of variables which includes the auxiliary fields
it can be proven that the field redefinition solving
the splitting problem is actually a canonical transformation.
We will deal with TYM in sect.~\ref{TYM}.

\subsection{A shortcut}    
    
Sometimes, there is an easy shortcut for finding the correct splitting      
functions. We notice that at first order $\Pi_{\phi,i} = \xi_i$ provides a solution to      
equations~(\ref{ns_4}). This suggests that if we are able to find coordinate transformations      
(which will  eventually be expressed by means of canonical transformations) such that      
the r.h.s. of (\ref{ns_0}) becomes linear in the quantum fields $\xi_i$, the solution    
$\Pi_{\phi,i} = \xi_i$  gives an all-order solution to the problem~(\ref{ns_4})    
and this allows us to identify the splitting.      
     
Let $\Phi' = \Phi'(\Phi)$ be a suitable change of coordinates such that      
\begin{eqnarray}\label{ns_1}     
s\, \Phi'_i = (R') ^\a_i[\Phi'] C_\a\,, ~~~~~~~~~~      
s\, C_\a =  {1\over 2} F_\a^{\beta\gamma}[ \Phi(\hat\phi,\xi)] C_\beta C_\gamma     
\end{eqnarray}     
and $(R') ^\a_i[\Phi'] $ is at most a linear function of $\Phi'$. Then,      
$\Pi_{\phi,i} = \xi_i$ is the trivial solution to the splitting problem, namely      
$\Phi' = \hat\phi' +   \xi_i$. Converting back to the original      
variables (again by means of a canonical transformation), which yield the      
inverted relation $ \Phi = \Phi(\Phi')$,      
we have      
\begin{equation}\label{gop1}     
\Phi_i = \Phi_i\left( \hat\phi' +   \xi_i \right) =  \Phi_i\left( \hat\phi'( \hat \phi) +   \xi_i \right) =      
\hat\phi_i + \Pi_i(\hat\phi,\xi)\,,      
\end{equation}     
where we substituted $\hat\phi'_i =  \hat\phi'_i( \hat \phi)$. 
Notice that to invert the change of coordinates      
we use the theorem for implicit functions in power series.      
     
In order to extend this analysis to more general theories, it is     
convenient to formulate the change of variables in the language of canonical      
transformations. The new set of fields and antifields are denoted by     
$\varphi^{' I}$ and $\varphi^{'*}_I$ ; they are related to the original      
variables by means of the transformation rules     
\begin{equation}\label{can1}     
\varphi^{' I} =(\Psi, \varphi^I)' = {\delta \Psi[\varphi,\varphi^{'*}] \over \delta \varphi^{'*}_I}\,,      
~~~~~     
\varphi^*_{I} =(\Psi, \varphi^*_I) = {\delta \Psi[\varphi,\varphi^{'*}] \over \delta \varphi^I}\,,      
\end{equation}     
where $\Psi$ is the generating functional of the canonical transformations.      
The bracket $(\cdot, \cdot)'$ is the bracket defined in (\ref{qn2_I}) with the      
coordinates $\varphi_I$ and $\varphi^*_I$ replaced by the new variables.      
The transformations of the new fields are computed using again      
the bracket     
\begin{eqnarray}\label{can2}     
s\, \varphi'_I = (S[\varphi', \varphi^{'*}], \varphi'_I)' \,, ~~~~     
s\, \varphi^{'*}_I = (S[\varphi', \varphi^{'*}], \varphi^{'*}_I)' \,,      
\end{eqnarray}     
and they extend the rules given in equation~(\ref{ns_1}).  
If the      
canonical transformations can be chosen in such a way that the new BRST      
transformation rules (\ref{can2}) are of the type given in equation~(\ref{ns_1})
(with $(R') ^\a_i[\varphi'] $ at most a linear function of $\varphi'$)
or completely linear in the quantum fields (as it happens
for the twisted version of N=2 SYM we will analyse 
in Sect.~\ref{TYM}),
we can split the fields and the antifields      
by      
\begin{eqnarray} \label{can3}      
\varphi'_I   =  \hat\varphi'_I  + \xi_I\,, ~~~~~      
 \varphi^{'*}_I =  \hat\varphi^{'*}_I + \xi^*_I\,,     
\end{eqnarray}     
where $\hat\varphi'_I$ and $\hat\varphi^{'*}_I $ are the background fields. The latter      
transform  according to (\ref{can2}) where $\varphi_I$ and $\varphi^*_I$ are replaced by      
the corresponding background fields. 
Notice that      
the canonical transformations do not need to be linear and in general they are      
analytical functions of the fields and antifields.

Even if the background fields $\varphi^*_I$  for      
the antifields are in principle not necessary, we found them useful as bookkeeping      
of the transformation rules for the background fields and they prove
to be convenient      
in order to formulate the canonical transformation in equation~(\ref{can1}).      
     
By an other (inverse) canonical transformation, we can re-express the new variables      
$\varphi'_I$ and $\varphi^{'*}_I $ in terms of the older ones: this leads to the relation      
between the original fields $\varphi_I$ and $\varphi^*_I$ and the      
quantum fields $\xi_I$ and $\xi^{*}_I$.

\subsection{A simple example}  
 \label{simple_example} 
  
As a warming-up example, we consider the simple topological Yang-Mills theory.   
This example is interesting because it displays some of the features of the N=2 model   
that will be discussed later, but at the same time is very simple. In the present   
example we will show how to construct the BFM by using 
the Jordan form of the BRST differential, which can be reached in the
present model by a simple field redefinition,
and  how to use the BFM to characterize the BRST cohomology and   
the physical observables of the theory.   
  
The observables of the theory   
are not defined in terms  of the BRST cohomology only, but a supplementary condition   
is needed. In fact, $H(s)$ and $H(s|d)$ are empty for any ghost number.   
This can be easily verified by using a suitable canonical transformation of variables which   
brings all the transformation rules into the form of trivial pairs ($s {\cal U} = {\cal V}$ and 
$s {\cal V} =0$). This is discussed in the next paragraphs. On the other side one can   
define a new nilpotent BRST-like operator $w$ associated with the gauge invariance and with   
the independence of the classical 
ungauged action from the ghost field $c$, such that   
the observables are identified by the cohomologies  
\begin{equation}\label{equivacoho}  
H_{basic}(s) = \Big\{ H(s| {\cal B }) | w {\cal B}  =0 \Big\} \,,  
~~~~~~  
H_{basic}(s|d) = \Big\{ H(s| {\cal B }) | w {\cal B} = d {\cal X} \Big\} \,.  
\end{equation}  
Here ${\cal B}$ is the space of basic forms which are gauge-invariant and   
do not depend on the ghost $c$. In the following we will construct the   
BRST-like operator $w$ and discuss the relation with the BFM. 
  
The topological Yang-Mills theory is described by the BRST transformations:      
\begin{eqnarray}     
s\, A = \psi - \nabla c \,, ~~~ s\, \psi =  [\psi, c] - \nabla\phi \,, ~~~~      
s\, c = \phi - \frac{1}{2} [c,c] \, ~~~ s\,, \phi  = [\phi, c] \, ,     
\label{e1}     
\end{eqnarray}     
for the fields $\varphi_I = \{A, c, \psi, \phi \}$ and      
\begin{eqnarray}     
&& s\, A^* =-[A^*,c] - [\psi^*, \phi] \,, \nonumber \\     
&& s\, \psi^* =  A^* + [\psi^*,c] \,,  \nonumber \\     
&& s\, c^*     = \nabla A^* + [\psi^*, \psi] + [\phi^*,\phi] + [c^*,c]\,, \nonumber \\     
&& s\, \phi^*  = c^* + \nabla \psi^* + [\phi^*, c] \, ,     
\label{e1.1}     
\end{eqnarray}     
for the antifields $\varphi^*_I = \{A^*, c^*, \psi^*, \phi^* \}$. Fields and antifields are      
forms with values in the Lie algebra of the underlying gauge group. The antifields $\varphi^*_I$      
are defined as the Hodge dual of the conventional definition, for example      
$A^* = A^{* \mu} \epsilon_{\mu\nu\rho\sigma} dx^\nu dx^\rho dx^\sigma$ is a $3$-form      
in $4$ dimensions.     
The background fields $\hat \varphi^I$ and $\hat\varphi^*_I$      
transform correspondingly, 
according to the considerations of the previous section.    
The fields $\hat\varphi^*_I$ are introduced in order      
the quantum fields $\xi_I$ to be coupled to their ``quantum'' antifields $\xi^{*I}$,    
after the splitting.     
      
It is easy to see that with the change of coordinates      
$\psi ' = \psi - \nabla c \,, \phi' = \phi - \frac{1}{2} [c,c]\,, A^{'*} = A^* + [\psi^*,c]$ and      
$c^{'*} = c^* + \nabla \psi^* + [\phi^*, c]$, generated by the functional      
\begin{equation}     
  \label{can4}     
\Psi[\varphi_I, \varphi^{'*}_I]  = \int {\rm tr} \left[A^{'*} A + c^{'*} c + \psi^{' *} \left( \psi - \nabla c  \right) +      
\phi^{'*} \left( \phi - \frac{1}{2} [c,c] \right) \right]\,,     
\end{equation}     
the BRST transformations (\ref{e1}) and  (\ref{e1.1}) become linear in the new variables.       
This leads to the consequence that the cohomologies $H(s)$ and $H(s|d)$ are empty.  
Then we can split the new fields in a linear way by setting     
\begin{eqnarray}     
&&      
A = \hat A + \xi_A \, , ~~~      
\psi' = \hat \psi' + \xi'_\psi \,, ~~~      
c = \hat c + \xi_c \, , ~~~      
\phi' = \hat \phi' + \xi'_\phi \,,      
\nonumber \\&&      
A^{'*} =  \hat A^{'*} + \xi'_{A^*} \,, ~~~      
\psi^* = \hat\psi^* + \xi_{\psi^*}\,, ~~~     
c^{'*} = \hat c^{'*} + \xi'_{c^*} \,, ~~~      
\phi^* = \hat \phi^* + \xi_{\phi^*} \,. 
\label{e3}     
\end{eqnarray}     
The BRST transformations of the new variables are the obvious ones     
derived from equations~(\ref{e1}) and (\ref{e1.1}),      
i.e. for instance in the case of the doublet $(A,\psi')$:     
$s \hat A = \hat \psi' \, , s\, \xi_A = \xi'_\psi \,, s \hat \psi' = 0$ and  $s \xi'_\psi = 0$.       
The old quantum fields are obtained by exploiting equation~(\ref{e1}).     
For instance, in the case of $\psi'$ we obtain     
\begin{eqnarray}     
\psi = (\hat\psi' + d \hat c + [\hat A,\hat c])  + \xi'_\psi + d\xi_c + [\hat A, \xi_c]      
           + [\xi_A,\hat c]  + [\xi_A, \xi_c] \, .     
\label{e5}     
\end{eqnarray}     
The terms in the round brackets are of order zero in the quantum fields,      
the remaining terms contain one or two powers of the quantum fields.     
Notice that the splitting in equation~(\ref{e5}) is non-trivial and non-linear      
in the quantum fields.\footnote{In several examples, one can use a superfield notation      
${\cal A} = c + A + F^* + \psi^* + \phi^*$ and ${\cal B} = \phi + \psi + F + A^* + c^*$. $F$ is the      
field strength and $F^*$ is the antifield associated with the condition of 
self-duality.      
The BRST transformations (\ref{e1}) and (\ref{e1.1}) can be written in a compact manner as      
\begin{equation}\label{foot1}     
 (s + d) \, {\cal A} + {\cal A}^2 = {\cal B}\,, ~~~~~ (s +d) \, {\cal B} + [{\cal A}, {\cal B}] =0 \nonumber\,.     
\end{equation}     
Thus, by introducing the new variables ${\cal B}' =  {\cal B} - {\cal A}^2$ and ${\cal A}' = {\cal A}$,      
the BRST transformations are simplified and the splitting becomes trivial.}     
     
To define the observables of the theory, we consider again the transformation rules given   
in (\ref{e1}) and those for the background fields  
\begin{eqnarray}     
s\, \hat A = \Omega - \hat\nabla \hat c \,, ~~~ s\, \Omega =  [\Omega, \hat c] - \hat\nabla\theta \,, ~~~~      
s\, \hat c = \theta - \frac{1}{2} [\hat c, \hat c] \, ~~~ s\, \theta  = [\theta, \hat c] \,.    
\label{e1.back}     
\end{eqnarray}     
Again, it is convenient to redefine the fields $\Omega' = \Omega - \hat\nabla \hat c$ and 
$\theta' = \theta - \frac{1}{2} [\hat c, \hat c]$ in order to simplify the relation between our notation and the one adopted in 
\cite{Stora_TYM}. The definition of the basic forms is obtained by computing the kernel of the 
operator
\begin{eqnarray}     
&&w\,  A = - \nabla \hat c \,, ~~~   
w\, \psi =  [\psi, \hat c] \,, ~~~~ \nonumber \\ 
&&w\,  c = \theta' + [c, \hat c] \, ~~~ 
w\, \phi  = [\phi, \hat c] \,, ~~~   
\nonumber \\   
&&w\, \hat c = - \frac{1}{2} [\hat c, \hat c] \, ~~~   
w\, \theta'  = [\theta', \hat c] \,.  
\label{e2.back}     
\end{eqnarray}     
As it can be readily seen the operator $w$ is nilpotent and anticommutes with   
the BRST symmetry (\ref{e1})-(\ref{e1.back}). This transformation rule can be extended 
in order to take into account the background field $\hat A$. We have in addition 
\begin{equation}\label{ee2.back}
w \, \hat A = - \hat \nabla \hat c\,, ~~~~~ w \, \Omega' = [\Omega', \hat c]\,.
\end{equation}
The transformation rules for the gauge field  $A$ and its partner $\hat A$ are the usual 
background gauge transformations. The   
rule for the background ghost $\hat c$ is the usual transformation 
for the gauge ghost. 
Notice that in addition the transformation for the ghost $\Omega'$ is the   
usual background trasformation. Finally, we have to point out that all the   
transformations given in (\ref{e2.back}) are linear in quantum   
fields and therefore they lead to linear WTI.   
  
Comparing (\ref{e2.back}) with the operator $w$ given in \cite{Stora_TYM}, one   
can see that all  transformations do coincide except those for the background   
$\hat A$ which are indeed new. The purpose of the operator $w$ 
is to restrict the  
space of local operators
to the sector of basic forms and  it is   
fundamental to define the observables at the quantum level. It happens that   
the construction of this operator in the BFM context is completely natural since the   
fields $\hat c$ and $\theta$, necessary to implement the gauge   
transformations, are indeed present. We can therefore conclude that the   
restriction to the space of those background 
gauge-invariant
 polynomials which are independent   
of the ghost $c$ (notice that this requirement 
is implemented by means of the linear   
shift into $\theta$) gives the correct observables. The BFM is not only a useful tool   
to compute gauge-invariant operators correlation functions, 
but it is also fundamental to select the physical content of the theory.   

\medskip
Finally, we can summarize the results of the present section in the following      
remarks. We find out that, according to a cohomological analysis or by use of       
suitable field redefinitions, we can derive the splitting functions $\Pi_\phi, \Pi_{\phi^*},      
\dots, \Pi_{C^*}$, such that the background gauge transformations of the      
quantum fields $\xi^I_{\phi}$ and $\xi_{C \a}$  are linear in the quantum fields.      
By defining the operator ${\cal N}_{\hat c} = \int \hat c_\a \delta_{\hat c_\a}$,       
which counts the powers of the background ghost fields $\hat c_\a$,      
we can decompose the BRST operator $s$ in terms of eigenvalues      
of ${\cal N}_{\hat c}$: $s = s_0 + s_1 + \sum_{n > 1} s_n$ where $s_0$ represents the      
BRST operator for the classical BRST symmetry, $s_1$ entails the background gauge invariance,      
$s_n$ with $n>1$ describe the closure terms.

\subsection{Background Gauge Fixing}\label{bgfix}
     
The splitting problem defines a  change of variables   
such that the new quantum fields transform linearly under the   
background transformations.    
If the splitting problem cannot be solved, the background transformations   
cannot be defined, independently of the perturbative quantization   
of the theory.    
In those cases where the good variables, suited for the implementation    
of the BFM, have been found, an additional problem arises: is it possible
to find a suitable gauge-fixing condition compatible with the invariance
of the ungauged classical action under the WT background
identities?
This issue can be analysed in a very general setting within   
the BV formalism and has already been thoroughly considered   
in the literature\footnote{For a review see e.g. Ref.~\cite{Gomis:1994he}.}.   
Here we only discuss some aspects relevant to the simplest case   
of irreducible models.

In order to construct the quantum effective action in perturbation theory,   
we need to compute the propagators for all quantum fields $\xi_{\phi_i}$ of   
the theory.    
The ungauged classical action $S$, fulfilling the master equation   
\begin{eqnarray}   
(S,S)=0 \, ,   
\label{gf0}   
\end{eqnarray}   
gives rise to a matrix of 2-point functions which is in general   
non-invertible. In order to remove this degeneracy, $S$ must be   
modified by adding non-minimal sectors. Then one  performs   
a canonical transformation, generated by the gauge-fixing fermion $\Psi_{g.f.}$,   
in such a way that the transformed classical action yields   
well-defined propagators for all quantum fields.   
   
The ungauged classical action $S$ depends on the background fields   
$\hat \phi_i$ and on the new quantum fields $\xi_{\phi_i}$. It fulfills   
the background WT identities, under which $\xi_{\phi_i}$ transform   
linearly.   
The addition of the non-minimal sectors, needed to fix the gauge,   
and the canonical transformation generated by the gauge-fixing   
fermion $\Psi_{g.f.}$ should not break this WT invariance.    
   
The minimal sectors we will analyse involve   
one generation of antighost fields $\bar c^\a$ and 
Lagrange multipliers $B^\a$,    
together with the corresponding antifields.      
The index $\a$ runs over those fields $\phi_\a$ whose
2-point function matrix $\{ \G_{\phi_\a \phi_\beta} =
\frac{\delta^2 S}{\delta \phi_\a \delta \phi_\beta} \}$,
computed from the ungauged classical action $S$, is not invertible.

As a first step, we add to the ungauged classical    
action $S$ the non-minimal terms      
\begin{equation}\label{gf1}     
S_{\rm n.m.} = \int  \bar c^{* \a} B_\a \, .   
\end{equation}     
Then we implement the gauge-fixing by means of a canonical transformation      
generated by the gauge-fixing fermion functional      
\begin{equation}\label{gf2}     
\Psi_{g.f.}[\hat\phi, \xi] = \int \bar c^\a {\cal F}_\a(\hat\phi_i, \xi_{\phi_i}, \hat c^\a, \xi_C \,; B^\a) \,.     
\end{equation}     
Let us denote by $\G^{(0)}$ the action obtained from $S+S_{n.m.}$
after the gauge-fixing canonical transformation has been performed.
The ${\cal F}_\alpha$ in equation~(\ref{gf2}) are chosen in such a way that,   
after the canonical transformation, the complete matrix of the 2-point functions,   
computed from $\G^{(0)}$, is invertible.   
   
The extension of the    
background transformations to the generators $(\bar c^\a, B^\a)$   
of the non-minimal sector must yield background transformations   
for  $(\bar c^\a, B^\a)$   
which are linear in the quantum fields.   
Moreover, we also require that 
the transformed gauge-fixed classical action $\G^{(0)}$
obeys the background WT invariance.
This requirement is fulfilled if we 
impose that the functional $\Psi_{g.f.}$ is background-gauge   
invariant:   
\begin{eqnarray}   
\delta_{bkg} \Psi_{g.f.} = 0 \, .   
\label{newgf1}   
\end{eqnarray}   
where $\delta_{bkg}$ denotes here the component of $s$ of degree one   
in the background ghost fields (the generator of the background transformations),    
properly extended to the non-minimal sector.   
From equation~(\ref{newgf1}) one gets   
\begin{eqnarray}   
\delta_{bkg} \bar c^\a {\cal F}^\a - \bar c^\a \delta_{bkg} {\cal F}^\a = 0 \, .   
\label{gfnew2}   
\end{eqnarray}   
By taking into account the above equation and the requirement   
of the linearity of $\delta_{bkg} \bar c^\a$ we obtain   
\begin{eqnarray}   
\delta_{bkg} \bar c^\a = {\cal M}^{\a \beta}(\hat \phi,\hat c) \bar c^{\beta} \, ,   
\label{gfnew4}   
\end{eqnarray}   
where ${\cal M}^{\a \beta}(\hat \phi,\hat c)$ is independent   
of the quantum fields $\xi_{\phi_i}, \bar c^\a,B^a$.   
Eq.(\ref{gfnew4}) provides the natural definition for the background transformation   
of $B^\a$:   
\begin{eqnarray}   
\delta_{bkg} B^\a = {\cal M}^{\a \beta}(\hat \phi,\hat c) B^\beta \, .   
\label{gfnew5}   
\end{eqnarray}    
By substituting eq.(\ref{gfnew4}) into eq.(\ref{gfnew2}) we get    
that the functions ${\cal F}_\a(\hat\phi_i, \xi_{\phi_i}, \hat c^\a, \xi_c \,; B^\a)$    
should transform as follows under $\delta$:   
\begin{eqnarray}   
\delta_{bkg} {\cal F}^\a = {\cal M}^{\beta \a} {\cal F}_\beta \, .   
\label{gfnew6}   
\end{eqnarray}

The fact that the fields      
$\xi_{\phi_i}$ and $\xi_C$ transform under linear background gauge transformations      
simplifies the construction of the functions ${\cal F}_\a$:    
it turns out that in many cases, as for instance in ordinary gauge   
theories,    
they can be obtained from their background-independent component   
 by covariantizing the differential operators with respect to the background fields.      
   
The case of TYM,  which we will analyse in Sect.~\ref{TYM}, is rather 
peculiar. There we first perform   
the gauge-fixing of the classical action in terms of the original   
unsplitted variables. The solution to the splitting problem yields 
for TYM a set of variables that transform  linearly under the full BRST   
differential, which can hence be identified with the generator   
of the background symmetry. As a consequence, the gauge-fixing term   
does not need to be modified to respect the background invariance.   
   
 Once a background covariant gauge-fixing has been introduced,   
the quantum effective action can be constructed in perturbation theory.   
The symmetry requirements of background invariance and ST invariance   
at the quantum level    
can be discussed along the lines of 
\cite{Grassi:1995wr,Grassi:1999nb,Becchi:1999ir,Ferrari:2000yp}.
   
As a final point, we would like to emphasize that the BFM construction   
of physical connected amplitudes requires the introduction of    
an additional gauge-fixing term for the classical background gauge   
fields \cite{Becchi:1999ir}. The latter does not affect the computation of the   
quantum effective action and only enters in the BFM computation    
of connected amplitudes of BRST-invariant local operators.   
A complete discussion of the interplay between this   
background gauge-fixing term, the background WT identities   
and the ST identities is provided in \cite{Becchi:1999ir,Ferrari:2000yp}.   
   
\section{N=2 Super Yang--Mills}\label{N2SYM}
     
\subsection{Topological Yang--Mills theory}  \label{TYM}   
     
In this section we show how the background field method can be implemented     
for $N=2$ super-Yang--Mills in the Wess--Zumino gauge. 
We will work within the flat Euclidean space-time.      
In order to construct the correct splitting of the fields 
into a background and a quantum part, with the latter transforming 
linearly under the background symmetry, we consider the off-shell 
formulation of the supersymmetry algebra of twisted $N=2$  
super-Yang-Mills in the Wess-Zumino (WZ) gauge. 
 
In the off-shell formulation the fields of $N=2$ super-Yang-Mills 
in the WZ gauge consist of a gauge field $A_\mu$, two spinors 
$\psi^i_\alpha$, $i=1,2$ and the conjugate $\bar \psi^i_{\dot{\alpha}}$, 
two scalars $\phi,\bar \phi$ ($\bar \phi$ being the complex 
conjugate of $\phi$) and an $SU(2)$ triplet of auxiliary fields 
$b^{ij} = b^{ji}$, $i,j=1,2$. 
 
After the twisting and the identification of the interal index  
$i$ with the spinor index $\alpha$, the spinor $\bar \psi^i_{\dot{\alpha}}$ 
can be related to an anticommuting vector $\psi_\mu$ given by 
\begin{eqnarray} 
\psi_\mu = (\bar \sigma_\mu)^{\dot{\alpha} \alpha} \bar \psi_{\alpha \dot{\alpha}} \, . 
\label{tw1} 
\end{eqnarray} 
The fields $\psi_{\alpha\beta}$ are decomposed into their 
symmetric component $\psi_{(\alpha\beta)}$ and  
their antisymmetric component $\psi_{[\alpha\beta]}$: 
\begin{eqnarray} 
\psi_{\alpha\beta}= \psi_{(\alpha\beta)} +  
                    \psi_{[\alpha\beta]} \, . 
\label{tw2} 
\end{eqnarray} 
$\psi_{(\alpha\beta)}$ is related to an antisymmetric self-dual 
anticommuting field $\chi_{\mu\nu}$ via the definition 
\begin{eqnarray} 
\chi_{\mu\nu} = \tilde \chi_{\mu\nu} = (\sigma_{\mu\nu})^{\alpha\beta} 
\psi_{(\alpha\beta)}  
\label{tw3} 
\end{eqnarray}  
where $\tilde \chi_{\mu\nu} = \frac{1}{2} \epsilon_{\mu\nu\rho\sigma} 
\chi^{\rho\sigma}$. 
 
The antisymmetric component $\psi_{[\alpha\beta]}$ is associated 
to the anticommuting scalar $\eta$ given by 
\begin{eqnarray} 
\eta= \epsilon^{\alpha\beta} \psi_{[\alpha\beta]} \, . 
\label{tw4} 
\end{eqnarray} 
Finally the auxiliary fields $b_{\alpha\beta}$ are related to 
the antisymmetric commuting self-dual field $b_{\mu\nu}$ 
defined by 
\begin{eqnarray} 
b_{\mu\nu} = (\sigma_{\mu\nu})^{\alpha\beta} b_{\alpha\beta} \, . 
\label{tw5} 
\end{eqnarray} 

Therefore the off-shell multiplet of $N=2$ super-Yang-Mills in the 
Wess-Zumino gauge $(A_\mu, \psi^i_\alpha, \bar \psi^i_{\dot{\alpha}}, 
\phi, \bar \phi, b^{ij})$ is transformed into the twisted 
multiplet  
$$(A_\mu, \psi_\mu, \chi_{\mu\nu}, \eta, \phi, \bar \phi, b_{\mu\nu}),$$ 
providing the field content of topological Yang-Mills theory (TYM) 
in the off-shell formulation. 
     
The classical action of TYM is given by     
\begin{eqnarray}     
S_{TYM} & = & \frac{1}{g^2} Tr \int d^4x \, \left (     
              +\frac{1}{2} F_{\mu\nu}^- F^{\mu\nu -}      
              -\frac{1}{2} b_{\mu\nu}b^{\mu\nu}    
              - \chi^{\mu\nu} (D_\mu \psi_\nu - D_\nu \psi_\mu)^- 
              \right . \nonumber \\     
        &   & \left . ~~~~~~~~ + \eta D_\mu \psi^\mu      
              - \frac{1}{2} \bar \phi D_\mu D^\mu \phi      
              + \frac{1}{2} \bar \phi \{ \psi^\mu, \psi_\mu \}      
              \right .     
              \nonumber \\     
        &   & \left . ~~~~~~~~ -\frac{1}{2} \phi      
              \left \{ \chi^{\mu\nu}, \chi_{\mu\nu} \right \}      
              - \frac{1}{8} [\phi,\eta] \eta      
              - \frac{1}{32} [\phi, \bar \phi] [\phi, \bar \phi]      
              \right ) \, .     
\label{ty1}     
\end{eqnarray}     
The action in equation~(\ref{ty1}) coincides with the one     
given in \cite{TYM_sorella} when the equation of motion     
for the auxiliary field $b_{\mu\nu}$ is imposed. $D_\mu$ is the covariant derivative given by      
$D_\mu (\cdot) =\partial_\mu \cdot +  [A_\mu,\cdot]$.    
 We denote by a $-$ the self-dual component of a tensor, so that     
\begin{eqnarray}     
F^-_{\mu\nu} = F_{\mu\nu} +      
\frac{1}{2} \epsilon_{\mu\nu\rho\sigma} F^{\rho\sigma} \, .     
\label{ty2}     
\end{eqnarray}     
$F^-_{\mu\nu}$ fulfills     
\begin{eqnarray}     
\tilde F^-_{\mu\nu} = \frac{1}{2} \epsilon_{\mu\nu\rho\sigma}    
F^{\rho\sigma-} = F^-_{\mu\nu} \, .     
\label{ty3}     
\end{eqnarray}     

As is well-known \cite{TYM_sorella}, the classical TYM action     
can be regarded as the twisted version of $N=2$ super-Yang--Mills theory     
in the Wess--Zumino gauge. As a consequence, in addition to gauge     
invariance, the classical TYM action exhibits further     
symmetries generated by      
the twisted $N=2$ supersymmetry generators \cite{TYM_sorella}.     
The set of these  generators contains     
a scalar generator $\delta$, a vector generator $\delta_\mu$ and     
a self-dual tensor generator $\delta_{\mu\nu}$, where      
$\delta$ is to be identified with Witten's fermionic symmetry   
\cite{witten_top}.

For our purposes we find it convenient to gather     
the BRST symmetry $s$, issued from gauge invariance     
of $S_{TYM}$ in equation~(\ref{ty1}),      
the scalar symmetry $\delta$ and the     
vector symmetry $\delta_\mu$, together with translation invariance,     
into a single BRST differential ${\cal Q}$ \cite{TYM_sorella},     
given by     
\begin{eqnarray}     
{\cal Q} = s + \omega \delta + \epsilon^\mu \delta_\mu + v^\mu \partial_\mu     
             - \epsilon^\mu \frac{\partial}{\partial v^\mu} \, .     
\label{ty4}       
\end{eqnarray}     
Here, $\omega$ is a commuting constant external source associated with     
Witten's fermionic symmetry.     
We remark that, unlike in \cite{TYM_sorella}, $\omega$ does not carry     
any ghost number. This is reflected in our assignment of the      
ghost number for the fields of the model:     
$\psi_\mu$ is assumed to carry ghost number $1$,     
$\phi$ ghost number $2$, $\chi_{\mu\nu}$ and $\eta$ ghost number $-1$,     
while $\bar \phi$ carries ghost number $-2$;    
$A_\mu$ and $b_{\mu\nu}$ carry zero ghost number.     
The constant external source     
 associated with the vector symmetry is denoted by $\epsilon^\mu$, the     
constant external source associated with translations by $v^\mu$.
With our assignments, $\epsilon_\mu$ carries ghost number 2,     
while $v^\mu$ carries ghost number 1.

As noted in \cite{TYM_sorella}, we can discard the tensor generator     
$\delta_{\mu\nu}$, since it does not carry additional information     
with respect to the subalgebra generated by $s$, $\delta$, $\delta_\mu$     
and $\partial_\mu$. 
The explicit form of the operator ${\cal Q}$ is given in Appendix \ref{TYM_BRST}.     
Since we are using the off-shell formalism with the auxiliary fields 
$b_{\mu\nu}$, ${\cal Q}^2=0$. 
In the on-shell formalism adopted e.g. in  
\cite{TYM_sorella}, where the auxiliary fields $b_{\mu\nu}$ are eliminated 
via their equation of motion, the 
operator ${\cal Q}$ is nilpotent only modulo the equations 
of motions of $\psi_\mu$ and $\chi_{\mu\nu}$; the corresponding STI can 
be written by adding suitable terms quadratic in the antifields
coupled to $\psi_\mu$ and $\chi_{\mu\nu}$.
     
We can now gauge-fix the classical TYM action by choosing \cite{TYM_sorella}:     
\begin{eqnarray}     
S_{gf} & = & {\cal Q} \int d^4x \, Tr \left ( \bar c \partial A \right )     
\nonumber \\     
& = & Tr \int d^4x \left (     
b\partial A + \bar c \partial^\mu D_\mu c - \omega \bar c \partial^\mu \psi_\mu     
- \frac{\epsilon^\nu}{2} \bar c \partial^\mu \chi_{\nu\mu}      
- \frac{\epsilon^\mu}{8} \bar c \partial_\mu \eta \right ) \, .     
\label{ty4_bis}     
\end{eqnarray}     
In the above equation $\bar c$ is the antighost field and $b$     
is the Nakanishi--Lautrup multiplier field.     
The gauge-fixed classical action      
\begin{eqnarray}     
\Sigma = S_{TYM} + S_{gf}      
\label{ty4_ter}     
\end{eqnarray}     
is ${\cal Q}$-invariant.     
     
In order to write the ST identities we couple the ${\cal Q}$-variations     
of the fields to the corresponding antifields as follows:     
\begin{eqnarray}     
S_{ext} & = &  Tr \int d^4x \, \left ( c^* {\cal Q}c + \phi^* {\cal Q}\phi      
+ A^{\mu *} {\cal Q} A_\mu + \psi^{\mu *} {\cal Q} \psi_\mu +\bar c^* {\cal Q} \bar c \right .     
\nonumber \\     
        &   & ~~~~~~~~ \left . + b^* {\cal Q}b + {\bar \phi}^* {\cal Q}\bar \phi + \eta^* {\cal Q}\eta     
                       + \frac{1}{2} \chi^{\mu\nu *} {\cal Q}\chi_{\mu\nu}      
                       + \frac{1}{2} b^{\mu\nu*} {\cal Q} b_{\mu\nu} \right ) \nonumber \\     
        &   & ~~~~~~~~ + v^{\mu*} {\cal Q} v_\mu  \, .     
\label{tyaf1}     
\end{eqnarray}     
The full classical action is then given by     
\begin{eqnarray}     
\G^{(0)} = S_{TYM} + S_{gf} + S_{ext}  
\label{tyaf3}     
\end{eqnarray}     
and fulfills the following ST identities:     
\begin{eqnarray}     
{\cal S}(\G^{(0)}) & = &      
Tr \int d^4x \left ( \frac{\delta\G^{(0)}}{\delta A^\mu} \frac{\delta\G^{(0)}}{\delta A^{\mu *}}     
+ \frac{\delta\G^{(0)}}{\delta \psi^{\mu *}} \frac{\delta\G^{(0)}}{\delta \psi^{\mu}}     
+ \frac{\delta\G^{(0)}}{\delta c^*} \frac{\delta\G^{(0)}}{\delta c}     
+ \frac{\delta\G^{(0)}}{\delta \phi^*} \frac{\delta\G^{(0)}}{\delta \phi}     
+ \frac{\delta\G^{(0)}}{\delta {\bar \phi}^*} \frac{\delta\G^{(0)}}{\delta \bar \phi} \right .     
\nonumber \\     
& &      
\left . ~~~~~~ +\frac{\delta\G^{(0)}}{\delta {\eta}^*} \frac{\delta\G^{(0)}}{\delta \eta}     
               +\frac{1}{2} \frac{\delta\G^{(0)}}{\delta \chi^{\mu\nu *}} \frac{\delta\G^{(0)}}{\delta \chi_{\mu\nu}}     
               +\frac{1}{2} \frac{\delta \G^{(0)}}{\delta b^{\mu\nu*}}\frac{\delta \G^{(0)}}{\delta b_{\mu\nu}}     
               + \frac{\delta \G^{(0)}}{\delta \bar c^*}\frac{\delta \G^{(0)}}{\delta \bar c} \right .     
\nonumber \\     
&&      
\left . ~~~~~~ + \frac{\delta\G^{(0)}}{\delta b^*} \frac{\delta\G^{(0)}}{\delta b} \right )      
               +\frac{\delta \G^{(0)}}{\delta v_\mu^*} \frac{\delta \G^{(0)}}{\delta v^\mu}     
\nonumber \\     
& = & \frac{1}{2} (\G^{(0)},\G^{(0)})      
= 0 \, ,     
\label{tysti}     
\end{eqnarray}     
where the bracket is defined as      
\begin{eqnarray}     
&&      
\!\!\!\!\!\!\!\!\!\!\!\!\!     
(X,Y)  =  \int d^4x \sum_I  \sigma_I \left ( \frac{\delta X}{\delta \Phi_I}\frac{\delta Y}{\delta \Phi_I^*}      
                             - (-1)^{(\epsilon_X+1)} \frac{\delta X}{\delta \Phi_I^*}\frac{\delta Y}{\delta \Phi_I} \right ) \, .     
\label{tybracket}     
\end{eqnarray}     
In the above equation,      
 $\Phi_I = \{ A_\mu, \psi_\mu, \chi_{\mu\nu}, b_{\mu\nu}, \eta, \bar \phi, c, \phi, \bar c,b, v_\mu\}$ and     
$\sigma_I =0$ for all fields but $\chi_{\mu\nu}, b_{\mu\nu}$, for which      
$\sigma_{\chi_{\mu\nu}} = \sigma_{b_{\mu\nu}} = \frac{1}{2}$. This factor is needed to take into account     
antisymmetry in the Lorentz indices of $\chi_{\mu\nu}, b_{\mu\nu}$.     
the term $\epsilon(X)$ stands for the statistics of $X$ ($\epsilon(X)=0$ if $X$ is a boson,      
$\epsilon(X)=1$ is $X$ is a fermion).     
     
We also introduce the linearized ST operator $\tilde {\cal Q}$, given by     
\begin{eqnarray}     
\tilde {\cal Q} & = & (\G^{(0)},\cdot) \, .     
\label{tymlinsti}     
\end{eqnarray}     
Now we redefine the fields as follows:     
\begin{eqnarray}     
&& \omega \psi'_\mu - \partial_\mu c      
\equiv \tilde {\cal Q}A_\mu = \omega \psi_\mu - \partial_\mu c     
+ \dots \, , \nonumber \\     
&& \omega  b'_{\sigma\tau} \equiv \tilde {\cal Q} \chi_{\sigma \tau} =      
\omega b_{\sigma\tau} + \dots      
\, , \nonumber \\     
&& 2 \omega \eta ' = \tilde {\cal Q} \bar \phi = 2 \omega \eta + \dots \, , \nonumber \\     
&& - \omega^2 \phi' = \tilde {\cal Q}c = - \omega^2 \phi + \dots \, , \nonumber \\     
&& b' = \tilde {\cal Q} \bar c = b + v^\mu \partial_\mu \bar c \, ,     
\label{ty5}     
\end{eqnarray}     
while we leave all other fields unchanged.     
Notice that this transformation is invertible.     
Apart from the fields $(A_\mu,\psi'_\mu)$ only $\tilde {\cal Q}$-doublets     
are now present. Notice that the transformation generated   
by $\tilde {\cal Q}$ in equation~(\ref{ty5}) is     
now linear in the new quantum fields.     
     
Explicitly we get:      
\begin{eqnarray}     
&& \psi'_\mu = \psi_\mu - \frac{1}{\omega}  [A_\mu,c]  +      
              \frac{\epsilon^\nu}{2\omega} \chi_{\nu\mu}      
              +\frac{\epsilon_\mu}{8\omega} \eta  +      
              \frac{v^\nu}{\omega} \partial_\nu A_\mu \, ,      
\nonumber \\     
&& b'_{\sigma\tau} = b_{\sigma \tau} + \frac{1}{\omega}     
                     \{ c, \chi_{\sigma \tau} \}      
                    + F^-_{\sigma\tau}  
                      +\frac{\epsilon^\mu}{8\omega}      
                     (\epsilon_{\mu\sigma\tau\nu} + g_{\mu\sigma}g_{\nu\tau}     
                      -g_{\mu\tau}g_{\nu\sigma}) D^\nu \bar \phi      
\nonumber \\      
&& ~~~~~~ + \frac{v^\nu}{\omega} \partial_\nu \chi_{\sigma \tau} \, ,      
\nonumber \\     
&& \eta' = \eta + \frac{1}{2\omega} [c, \bar \phi] + \frac{1}{2\omega} v^\nu \partial_\nu \bar \phi  \, ,      
\nonumber \\     
&& \phi' = \phi - \frac{1}{\omega^2}      
c^2 + \frac{\epsilon^\mu}{\omega^2} A_\mu - \frac{\epsilon^2}{16\omega^2}      
      \bar \phi     
           + \frac{v^\nu}{\omega^2} \partial_\nu c \, , \nonumber \\     
&& b' = b + v^\mu \partial_\mu \bar c \, .     
\label{ty6}     
\end{eqnarray}     
The role played by $\epsilon_\mu$ and $v_\mu$ is rather     
suggestive: they can be thought as background fields entering     
into the field redefinition. From the cohomological point of view      
this is confirmed by the fact that $(v^\mu,\epsilon^\mu)$      
form a set of doublets under $\tilde {\cal Q}$. 
We remark that the field redefinition in equation~(\ref{ty5}) gives  
terms that are not analytic in $\omega$. We will discuss in 
Sect.~\ref{sec_equiv} how the BFM allows naturally to recover  
the observables of the model by taking into account the relevant 
equivariant cohomology of TYM. 
 
At this point we can perform a linear splitting in the primed variables     
\begin{eqnarray}     
\psi'_\mu = \hat \psi_\mu + \xi_{\psi_\mu} \, , ~~~~     
b'_{\sigma\tau} = \hat b_{\sigma\tau} + \xi_{ b_{\sigma\tau}} \, , ~~~~     
\eta' = \hat \eta + \xi_{\eta} \, ,  ~~~~     
\phi' = \hat \phi + \xi_{\phi} \, , ~~~~     
b' = \hat b + \xi_b \, ,\nonumber  \\     
\label{ty7}     
\end{eqnarray}     
and then go back to reconstruct the full non-linear splitting, 
making use of equation~(\ref{ty6}).      
Notice that also the fields     
that are unchanged under the field redefinition in equation~(\ref{ty5})     
are understood to be splitted into a background and a quantum part:     
\begin{eqnarray}     
& A_\mu = \hat A_\mu + \xi_{A,\mu} \, , ~~~     
\chi_{\sigma\tau} = \hat \chi_{\sigma\tau} + \xi_{\chi,\sigma\tau} \, , ~~~     
\bar \phi = \hat {\bar \phi} + \xi_{\bar \phi} \, , & \nonumber \\     
& c = \hat c + \xi_c \, , ~~~ \bar c = \hat {\bar c} + \xi_{\bar c}  \, . &     
\label{ty_split}     
\end{eqnarray}     

The corresponding BRST transformations of the new variables are     
given by     
\begin{eqnarray}     
&& \tilde {\cal Q} \hat A_\mu = \omega \hat \psi_\mu - \partial_\mu \hat c     
\, , ~~~~     
\tilde {\cal Q} \hat \psi_\mu = -\omega \partial_\mu \hat \phi \, , \nonumber \\     
&&     
\tilde {\cal Q} \xi_{A,\mu} = \omega \xi_{\psi_\mu} - \partial_\mu \xi_c      
\, , ~~~~      
\tilde {\cal Q} \xi_{\psi_\mu} = -\omega \partial_\mu \xi_\phi \, ,     
\label{ty7_bis}     
\end{eqnarray}     
\begin{eqnarray}     
\tilde {\cal Q} \hat \chi_{\sigma\tau} =  \hat b_{\sigma\tau} \, ,      
~~~~     
\tilde {\cal Q} \hat b_{\sigma\tau} = 0 \, ,      
~~~~     
\tilde {\cal Q} \xi_{\chi,\sigma\tau} =  \xi_{b,\sigma\tau} \, ,      
~~~~     
\tilde {\cal Q} \xi_{b,\sigma\tau} = 0 \, ,     
\label{tym_dopp}     
\end{eqnarray}     
and analogously for the other sets of $\tilde {\cal Q}$-doublets.     
     
As an example, in the case of $\psi_\mu$ we get      
\begin{eqnarray}     
\psi_\mu & = & \hat \psi_\mu + \xi_{\psi_\mu}     
               +\frac{1}{\omega} [\hat A_\mu, \hat c]     
               +\frac{1}{\omega} [\hat A_\mu, \xi_c]     
               +\frac{1}{\omega} [\xi_{A,\mu}, \hat c]     
               + \frac{1}{\omega}[\xi_{A, \mu}, \xi_c] \nonumber \\     
         &   & -\frac{\epsilon^\nu}{2\omega} ( \hat \chi_{\nu\mu} +      
                 \xi_{\chi, \mu\nu} )     
               - \frac{v^\nu}{\omega}     
                 \partial_\nu (\hat A_\mu + \xi_{A, \mu} ) \nonumber \\     
         &   & - \frac{\epsilon_\mu}{8\omega} \left ( \bar \eta + \xi_\eta - \frac{1}{2\omega} [\hat c, \hat{\bar\phi} ]     
               - \frac{1}{2\omega} [\hat c, \xi_{\bar \phi} ] \right . \nonumber \\     
         &   & \left . -\frac{1}{2\omega} [\xi_c, \hat{\bar \phi}] - \frac{1}{2\omega} [\xi_c, \xi_{\bar \phi}]      
               - \frac{1}{2\omega} v^\nu \partial_\nu (\hat{\bar \phi} + \xi_{\bar \phi} ) \right ) \, .     
\label{ty8}     
\end{eqnarray}     
Note that this splitting contains terms that are non-linear in the quantum fields.     
     
We recover the original $\tilde{\cal Q}$ transformation of $\psi_\mu$     
by acting with the $\tilde{\cal Q}$ transformations     
in equation~(\ref{ty7_bis}) on the R.H.S. of equation~(\ref{ty8}).     

We remark that, since the change of variables in equation~(\ref{ty5}) 
only involves fields, it is automatically a canonical transformation
in the space spanned by the fields and the antifields of the model. 
This is analogous to the previous example of TYM. 
 Therefore we do not modify the cohomology of the model 
while implementing the background splitting. 
In Sect.~\ref{sec_equiv} we show how to recover the relevant  
equivariant cohomology from the BFM.

A comment on the gauge-fixing function for TYM is in order.     
In the case of TYM we have been able to prove that the original BRST symmetry can be linearized     
by a suitable change of variables. This change of variables can be implemented via a canonical     
transformation, thus leaving the cohomology invariant.     
As a consequence, the classical ST identities in equation~(\ref{tysti}) hold. Moreover, they are already linear in the     
quantum fields, when expressed in terms of the new variables. We remark that these identities     
are fulfilled by the classical action whose gauge-fixing condition is the one given in equation~(\ref{ty4_bis}).     
No special choice of the gauge-fixing function is needed in the present case, since     
the full BRST symmetry becomes linear. This should be compared with the different     
situation in ordinary Yang-Mills theory (see Sect.~\ref{nonlinspl}).    
In these case the full BRST transformation cannot be cast in a linear form (see equation~({\ref{q3})), but  
one can establish an additional background WT identity,      
provided that a suitable background-dependent choice of the gauge-fixing function is made, as discussed  in Sect.~\ref{bgfix}.     
   
  With the conventions of \cite{TYM_sorella} on supersymmetry in Euclidean   
space-time we can go back to the original model N=2 SYM by using the  
map 
\begin{eqnarray}   
&& \psi_{(\alpha \beta)} = \frac{1}{4} (\sigma^{\mu\nu})_{\alpha\beta}   
\chi_{\mu\nu} \, , 
~~~~~~ \psi_{[\alpha \beta]} = \frac{1}{2} \epsilon_{\alpha\beta} \eta \, ,    
\nonumber \\   
&& \bar \psi_{\alpha \dot \alpha} = - \frac{1}{2}    
(\sigma^\mu)_{\alpha \dot \alpha} \psi_\mu \, ,  
~~~~~ b_{\alpha\beta} = \frac{1}{4} (\sigma^{\mu\nu})_{\alpha\beta} b_{\mu\nu} 
\, , 
\label{back1}   
\end{eqnarray}   
while the fields $A_\mu, \phi, \bar \phi$ are mapped into themselves.   
Due to the linearity of this map, the correct background splitting  
of the original N=2 multiplet in the WZ gauge can be directly recovered 
from the splitting of the twisted multiplet of TYM.

\subsection{Equivariant cohomology for TYM and the BFM} \label{sec_equiv}  
 
We showed that in order to introduce  
the BFM for N=2 SYM, it is convenient to perform the  twisting of the fields  
and to rewrite the theory as a topological model.  
By the canonical change of variables in equation~(\ref{ty5}) the BRST 
operator ${\cal Q}$ can be cast in the form 
${\cal Q} = \int d^4x {\cal V} \delta / \delta {\cal U}$, hence the cohomology 
of ${\cal Q}$ in the space of local formal power series 
vanishes. 
In addition, we know  
that N=2 SYM has a physical set of observables whose correlation functions  
do not vanish. 
Hence  
the observables of the theory should be defined  
not as the BRST cohomology on the entire space, but the latter should be restricted to a suitable subspace.  
Following  
\cite{Stora_TYM}, the correct set of observables is given by the BRST cohomology  
computed in the space of gauge invariant polynomials which are independent of the gauge ghost (in the  
literature this space is denoted as the space of {\it basic forms}).   
Within the BFM the observables are defined as the {\it BRST cohomology computed in the  
space of background gauge invariant polynomials which are independent of the
gauge ghost $c$} .  
This suggests that there exists a new nilpotent BRST operator $w$, associated with the background gauge  
symmetry (as in the example in Sect.~\ref{simple_example}), which permits to select the space of basic forms. 
 
As has been discussed in the previous section, we have to give up the analyticity in the  
ghost $\omega$ in order to implement consistently the splitting and the background  
invariance (with respect to the full BRST transformation generated by ${\cal Q}$) of the theory. 
However, the analiticity in $\omega$ turned to be a crucial ingredient in the analysis  
performed in \cite{TYM_sorella}. There, it has been shown how the request of analyticity allows to select the  
correct equivariant cohomology. Moreover, it can be proven  that by introducing a suitable  
grading of the fields in the theory the equivariant cohomology can always be selected by the space of polynomials  
in $\omega$. We cannot impose the analyticity requirement in order to identify the correct subspace, but 
we can construct a new differential $w$ whose kernel identifies the basic forms.  
 The differential ${w}$ is associated to the background gauge symmetry of the theory, therefore we neglect  
for the moment the contributions due to the supersymmetry and translations by setting $\epsilon^{\mu} = v^{\mu} =0$.  
 We will also drop the non-minimal doublets $(\bar c, b)$ since they are cohomologically trivial.  

The powers of $\omega$ entering the ${\cal Q}$-transformations  
induce a grading under which we can decompose ${\cal Q}$ as a sum of three terms:  
\begin{eqnarray}  
{\cal Q} = \tq_{-1} + \omega \tq_0 + \omega^2 \tq_1 \, .  
\label{eqcohomo3}  
\end{eqnarray} 
Explicitly, we have 
\begin{eqnarray}     
&& \tq_{0} A_\mu     =   \psi_\mu  \,, ~~~~~\tq_{0} \psi_\mu = - D_\mu \phi \,, \nonumber \\     
&& \tq_{0} \chi_{\sigma\tau}  =  F^-_{\sigma\tau} + b_{\sigma\tau}  \,,   ~~~~~~ \nonumber \\
&& \tq_{0} b_{\mu\nu} =  - (D_\mu \psi_\nu - D_\nu \psi_\mu)^- + [\chi_{\mu\nu}, \phi] \,,  \\ 
&& \tq_{0} \eta  =  \frac{1 }{2} [\phi,\bar\phi] \,, ~~~~~ \tq_{0} \phi  =  0 \, , \nonumber \\     
&& \tq_{0} \bar \phi  =  2  \eta  \, , ~~~~~~ \tq_{0} c = 0 \,. \nonumber 
\label{brsttqo}     
\end{eqnarray}    
It is  convenient to introduce $\tilde b_{\sigma\tau} = 
F^-_{\sigma\tau} + b_{\sigma\tau}$ and $\bar\phi  \rightarrow  \frac{\bar \phi}{2}$ to 
simplify these transformation rules. Now it is clear that $\tq_{0}$ is the de Rham operator and 
  $(\tq_{0})^{2} = {\cal L}_{\phi}$ where $ {\cal L}_{\phi}$ is the Lie derivative generated by 
the field $\phi$. The cohomology that we are looking for is the de Rham cohomology on the 
space of polynomials which are gauge invariant and independent of the ghost $c$. 
This means that the operator $w$ we are looking for is given by 
\begin{eqnarray}     
&& w A_\mu =  - D_\mu \hat c  \,,    ~~~~~~~ w \psi_\mu =  \{ \hat c, \psi_\mu \} \,,  \nonumber \\     
&& w \chi_{\sigma\tau}  =  \{ \hat c, \chi_{\sigma\tau} \} \,, ~~~~~~~  w \tilde b_{\mu\nu} = [\hat c, \tilde b_{\mu\nu}] \,, \nonumber \\ 
&& w \eta  =   \{ \hat c, \eta \} \,,   ~~~~~  w \phi  =   [\hat c,\phi]  \, , \nonumber \\     
&& w \bar \phi  =   [\hat c, \bar \phi]  \, ,  ~~~~~ w c  =  \{ \hat c, c\} - \hat \phi \, , \nonumber \\    
&& w \hat c  =  \hat c^2  \, , ~~~~~ w \hat \phi  =   [\hat c, \hat \phi]  \, ,
\label{ww1}     
\end{eqnarray}     
 The fields $\hat c$ and $\hat \phi$ correspond to the background ghost field and to the 
 background of $\phi$ respectively. 
Their BRST transformation are the usual contractible pair transformation
 \begin{equation}
 \tq_1 \hat c = - \hat \phi\,, ~~~~~~~  \tq_1 \hat \phi = 0\,.
 \label{ww2}
 \end{equation} 
As explained in the previous section, we also have all the background fields present. They 
transform under the BRST symmetry in the standard way, and under $w$ 
as a gauge transformation generated by $\hat c$:
\begin{eqnarray}
&& w \hat A_\mu =  - \hat D_\mu \hat c  \,,    ~~~~~~~ w \hat \psi_\mu =  \{ \hat c, \hat \psi_\mu \} \,,  \nonumber \\     
&& w \hat \chi_{\sigma\tau}  =  \{ \hat c, \hat \chi_{\sigma\tau} \} \,, ~~~~~~~  
w \tilde {\hat b}_{\mu\nu} = [\hat c, \tilde {\hat b}_{\mu\nu}] \,, \nonumber \\ 
&& w \hat \eta  =   \{ \hat c, \hat \eta \} \,,   ~~~~~  w \hat{\bar{\phi}}  =   [\hat c,\hat{\bar\phi}]  
 \label{ww3}
 \end{eqnarray}
The operator $w$ generates the gauge transformations of the model. Being linear in the quantum 
fields, it is possible to write linear WTI. We notice that by imposing the BRST invariance, and being 
all the background fields cohomologically trivial, it turns out that the observables will 
depend only upon the original fields. Finally, the background gauge invariance,
as expressed by equations.~(\ref{ww1}) and 
(\ref{ww3}), selects the gauge invariant observables. 
The linear shift of the background gauge 
transformation of $c$ implies that a gauge invariant operator annihilated by $w$ is independent 
of $c$. According to \cite{Stora_TYM}, the basic forms are identified with those  
local integrated functionals belonging to the kernel of $w$. 
 
The analysis of the cohomology follows the discussion of \cite{Ta-So}. 
We only point out that the generators of the equivariant  
cohomology classes are given by the polynomials  
\begin{equation}\label{cicAB} 
P(\phi) = \sum_n c_n Tr (\hat\phi + \xi_\phi)^{2n} 
\end{equation} 
where $c_n$ are numeric coefficients and the field $\phi$ is split into the quantum and background part. 
The combination $\hat\phi + \xi_\phi$ is fixed by the BRST symmetry and by the background gauge symmetry.  
 
In addition, we have that these  
polynomials are not BRST exact, indeed if they were 
they would have the form  
\begin{equation}\label{cicAA} 
P(\phi) = \sum_n c_n {\cal Q} \left( Tr (\hat c + \xi_c)(\hat\phi + \xi_\phi)^{2n-1} + \dots \right)\,, 
\end{equation} 
but $Tr (\hat c + \xi_c)(\hat\phi + \xi_\phi)^{2n-1} + \dots$  do not belong to the kernel  
of $w$ since $\hat c$ transforms into $\hat \phi$.  
 
Following the discussion in \cite{Ta-So}, there are further cohomological classes that  
are not eliminated by the previous argument, based on the background gauge invariance, for instance the operator:
\begin{equation} 
\Delta_{\mu\nu} = \left( F^-_{\mu\nu} +  b_{\mu\nu}  \right) \phi \,. 
\end{equation} 
 
This operator is not related to any observables of the N=2 SYM theory and therefore it should be absent  
from the cohomology. In \cite{Ta-So} 
it is excluded by imposing the supersymmetry with $\epsilon^{\mu}$. 
In our framework, by taking into account
the complete background symmetry for the complete differential ${\cal Q}$ instead of  
its gauge part, we found that the invariance under ${\cal Q}$ 
(which now contains the supersymmetry in its twisted version)  
excludes $\Delta_{\mu\nu}$.  
 
\newpage

\section{Conclusions and Outlook} 
 
We have discussed the implementation of the background field method from  
the geometrical point of view. 
Within the BV formalism
we have shown how the method should be generalized so as to deal 
 with open algebras and field-dependent structure constants. 
This requires the identification of the proper space of
variables on which the BFM splitting
problem can be defined.
In addition, we have underlined
the similarity between the background field method and the BRST symmetry for topological  
gauge theories. Using this idea, we have been able 
to formulate the BFM for N=2 SYM by introducing a  
field redefinition that brings the model in its topological twisted version.
Therefore, the required field splitting can be implemented 
 by a canonical  
transformation.
We have analysed the compatibility of this field redefinition
with the gauge-fixing procedure.
Finally we have shown
 that the BRST symmetry plus the background symmetry (which is now  
extended to all the symmetries of the model)
lead to the correct equivariant cohomology, needed to define the proper set 
of observables in  N=2 SYM.
These results should be regarded as a  step towards the construction
of a super BFM for the MSSM.  

\section*{Acknowledgements}     
We would like to thank R. Stora, P. van Nieuwenhuizen and M. Ro\v cek for      
useful discussions and comments. A.Q. and T.H. would like to thank 
Milan University, and  P.A.G. CERN and      
Milan University for the kind
hospitality. The research of P.A.G. is supported by the  
Grant PHY-0098526. 
    
\vfill 
 
\appendix     
\section{BRST transformations for TYM} 
\label{TYM_BRST}     
The operator ${\cal Q}$ in equation~(\ref{ty4}) acts as follows on the fields of TYM:     
\begin{eqnarray}     
&& {\cal Q} A_\mu =  -D_\mu c + \omega      
\psi_\mu + \frac{\epsilon^\nu}{2} \chi_{\nu\mu}     
                     +\frac{\epsilon_\mu}{8} \eta + v^\nu \partial_\nu A_\mu \, ,      
\nonumber \\     
&& {\cal Q} \psi_\mu =  \{ c, \psi_\mu \} - \omega D_\mu \phi +      
                        \epsilon^\nu \left ( F_{\nu\mu} - \frac{1}{2} F^-_{\nu\mu} - \frac{1}{4} b_{\nu\mu} \right )     
                       -\frac{\epsilon_\mu}{16} [\phi,\bar\phi] \nonumber \\     
                  && ~~~~~~~~~~~~~~~~ + v^\nu \partial_\nu \psi_\mu \, , \nonumber \\     
&& {\cal Q} \chi_{\sigma\tau}  =  \{ c, \chi_{\sigma\tau} \} +      
\omega F^-_{\sigma\tau} + \omega b_{\sigma\tau}      
                                 +\frac{\epsilon^\mu}{8} \left (\epsilon_{\mu\sigma\tau\nu}     
                                  +g_{\mu\sigma}g_{\nu\tau} - g_{\mu\tau}g_{\nu\sigma} \right )     
                                  D^\nu \bar \phi \nonumber \\     
                  && ~~~~~~~~~~~~~~~~           +v^\nu \partial_\nu \chi_{\sigma \tau} \, , \nonumber \\     
&& {\cal Q} b_{\mu\nu} = [c,b_{\mu\nu}] +  
                         \omega  ( - (D_\mu \psi_\nu - D_\nu \psi_\mu)^- 
                                    + [\chi_{\mu\nu}, \phi]) \nonumber \\ 
&& ~~~~~~~~~~~~~~~~ - \Big (  \epsilon_\mu (D^\tau \chi_{\nu\tau} 
                                            -D^\tau \chi_{\tau\nu} 
                            +\epsilon_{\nu\gamma\lambda\tau}  
                             D^\gamma \chi^{\lambda\tau}) + 
                              \epsilon_\mu D_\nu \eta - 
                              \epsilon_\mu [\psi_\nu,\bar \phi] \Big )^- 
                         \nonumber \\ 
&& ~~~~~~~~~~~~~~~~   + v^\rho \partial_\rho b_{\mu\nu} \, , \nonumber \\ 
&& {\cal Q} \eta  =   \{ c, \eta \} + \frac{\omega }{2} [\phi,\bar\phi] +      
                    \frac{\epsilon^\mu}{2} D_\mu \bar \phi + v^\nu \partial_\nu \eta \, , \nonumber \\     
&& {\cal Q} \phi  =   [c,\phi] - \epsilon^\mu \psi_\mu + v^\nu \partial_\nu \phi \, , \nonumber \\     
&& {\cal Q} \bar \phi  =   [c,\bar \phi] + 2\omega \eta + v^\nu \partial_\nu \bar \phi \, , \nonumber \\     
&& {\cal Q} c  =  c^2 - \omega^2 \phi - \epsilon^\mu A_\mu      
                        +\frac{\epsilon^2}{16} \bar \phi + v^\nu \partial_\nu c \, , \nonumber \\     
&& {\cal Q} v^\mu  =  - \omega \epsilon^\mu \, , \nonumber \\     
&& {\cal Q} \epsilon^\mu  =  0 \, , \nonumber \\     
&& {\cal Q} \omega = 0 \, , \nonumber \\     
&& {\cal Q} \bar c  =  b + v^\mu \partial_\mu \bar c \, , \nonumber \\     
&& {\cal Q} b  =  \omega \epsilon^\mu \partial_\mu \bar c + v^\mu \partial_\mu b \, .     
\label{brst1}     
\end{eqnarray}

   
\end{document}